\documentclass[a4paper,11pt]{article}
\usepackage{jheppub}

\usepackage{amsmath, latexsym, amssymb}
\usepackage{mathtools}
\usepackage{bm}
\usepackage{slashed}

\usepackage[T1]{fontenc}
\usepackage{xspace}

\usepackage{graphicx}
\usepackage{subcaption}

\usepackage{multirow}
\usepackage{makecell}
\usepackage{diagbox}
\usepackage{tabularx}
\usepackage{booktabs}
\usepackage{colortbl}

\usepackage{xcolor}
\definecolor{Gray}{gray}{0.9}
\definecolor{LightCyan}{rgb}{0.88,1,1}
\definecolor{nicered}{rgb}{.7,.1,.1}
\definecolor{nicegreen}{rgb}{.1,.5,.1}
\definecolor{darkblue}{rgb}{0,0,.5}
\hypersetup{colorlinks, citecolor=nicegreen, linkcolor=nicered, urlcolor=darkblue}

\usepackage{soul}
\usepackage{enumitem}
\usepackage{verbatim}
\usepackage{bbding, pifont}
\usepackage{academicons}
\usepackage{catchfilebetweentags}

\usepackage{siunitx}
\usepackage[capitalize]{cleveref}

\renewcommand\arraystretch{1.5}

\providecommand{\bbbar}{\mathrm{b}\overline{\mathrm{b}}}

\newcommand{\epem}{e^+e^-}

\newcommand{\sqrts}{\sqrt{s}}

\mathchardef\mhyphen="2D

\newcommand{\LumiInt}{\mathcal{L}_{\mathrm{\tiny{int}}}}

\newcommand*{\eg}{e.g.,\@\xspace}
\newcommand*{\ie}{i.e.,\@\xspace}
\newcommand*{\cm}{c.m.\@\xspace}

\newcommand{\orcid}[1]{\href{https://orcid.org/#1}{\hspace*{0.1em}\raisebox{-0.45ex}{\includegraphics[width=1em]{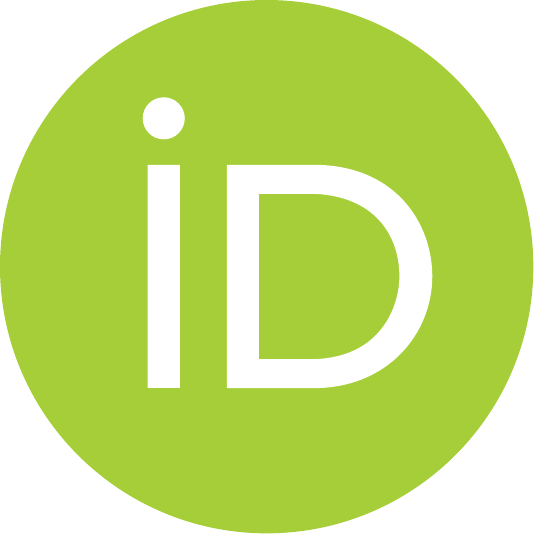}}}}

\title{Probing the electron Yukawa coupling via resonant Higgs boson production 
at FCC-ee via \texorpdfstring{$e^+e^- \to H \to WW^*$}{e+e- -> H -> WW*}
in lepton-plus-jets final states}

\author[a,b,c,d]{Apranik~Fatehi,\note{apranik.fatehi@cern.ch}}
\author[b,a]{Reza~Jafari Seyedabad,\note{reza.jafari@cern.ch}}
\author[b,e]{Amir~Amiri,\note{amir.amiri1308@gmail.com}}
\author[b,f]{Kazem~Azizi,\note{kazem.azizi@cern.ch}}
\author[a]{David~d'Enterria,\note{david.d'enterria@cern.ch}}
\author[g]{Louis~Portales,\note{louis.portales@cea.fr}}
\author[a]{and Michele~Selvaggi\note{michele.selvaggi@cern.ch}}

\affiliation[a]{CERN, EP Department,
CH-1211 Geneva, Switzerland}
\affiliation[b]{Department of Physics, University of Tehran,
North Karegar Avenue, Tehran 14395-547, Iran}
\affiliation[c]{Universit\"at Hamburg,
Luruper Chaussee 149, 22761 Hamburg, Germany}
\affiliation[d]{Deutsches Elektronen-Synchrotron DESY,
Notkestr. 85, 22607 Hamburg, Germany}
\affiliation[e]{Department of Physics, Faculty of Science, Ferdowsi University of Mashhad,
P.O.~Box 1436, Mashhad, Iran}
\affiliation[f]{Department of Physics, Dogus University,
Dudullu-\"{U}mraniye, 34775 Istanbul, T\"{u}rkiye}
\affiliation[g]{IRFU, CEA, Universit\'e Paris-Saclay,
Gif-sur-Yvette, France}

\abstract{We report a detailed simulation study of the search for $s$-channel Higgs boson production in \texorpdfstring{$e^+e^-$}{e+e-} collisions at a center-of-mass (\cm) energy of \texorpdfstring{$\sqrt{s}=125\,\mathrm{GeV}$}{sqrt(s)=125 GeV} at the CERN Future Circular Collider (FCC-ee), as a means to constrain the electron Yukawa coupling, $y_e$. The process of interest is $\epem\to H\to WW^*\to \ell^\pm\nu + jj$ with four different $WW^*$ final states considered, involving both on- and off-shell $W$ bosons decaying either into dileptons ($\ell^\pm = e^\pm$ and $\mu^\pm$, including those from $\tau^\pm$ decays) or into dijets ($jj$). Signal and background events are discriminated through a multiclass gradient boosted decision tree exploiting a comprehensive set of kinematic and topological variables across the four final-state categories. Assuming a monochromatized \cm\ energy spread of 4.1\,MeV, yielding a $\sigma_{ee\to H} = 280$\,ab resonant cross section, and an integrated luminosity of 10\,ab$^{-1}$, the analysis achieves a combined statistical significance of 2.0 standard deviations. This corresponds to an upper limit on the coupling modifier $\kappa_e = y_e/y_e^{\rm \textsc{sm}}\lesssim 1.35$ at 95\% confidence level, and provides the most stringent constraint on the electron Yukawa coupling achieved in simulation-based studies to date.}

\begin{document}
\maketitle
\flushbottom

\clearpage
\section{Introduction}

Following the observation at the Large Hadron Collider (LHC) of a scalar particle with a mass of $m_H \approx 125$\,GeV~\cite{Aad:2012tfa,Chatrchyan:2012ufa} consistent with the Standard Model (SM) Higgs boson, thereby confirming the SM electroweak symmetry breaking (EWSB) mechanism and the generation of elementary particle masses~\cite{Higgs:1964ia,Higgs:1964pj,Englert:1964et,Guralnik:1964eu}, the focus of Higgs physics has shifted from discovery to precision measurements of its properties. 
In the SM, fermion masses arise from Yukawa interactions with the Higgs doublet, where EWSB sets the proportionality between masses and their corresponding Yukawa couplings, leading to a striking hierarchy among fermion generations.
Apart from neutrinos, whose mass generation mechanism remains unknown, the smallest fermion Yukawa coupling is that of the electron, $y_e = \sqrt{2} m_e/v = 2.9\times10^{-6}$ given by the minuscule ratio of the electron mass ($m_e$) to the Higgs vacuum expectation value $v = (\sqrt{2}\mathrm{G_F})^{-1/2} = 246.22$\,GeV, and 
its measurement would provide a fundamental test of the SM~\cite{dEnterria:2021xij}. To date, only the Higgs couplings to the heavy gauge bosons and third-generation fermions have been directly confirmed experimentally at the LHC, establishing remarkable consistency with the SM predictions~\cite{Dittmaier:2012nh,Bass:2021acr,CMS:2022dwd,ATLAS:2022vkf}. The mass generation mechanism for the second-generation fermions will be only partially probed at the High-Luminosity LHC (HL-LHC)~\cite{Cepeda:2019klc,ATLAS:2025eii}. A direct measurement of $y_e$ would thus provide a powerful validation of the Higgs mass mechanism across all fermion generations.

The smallness of the electron-Higgs coupling renders it effectively unmeasurable in $pp$ collisions at the LHC, due to both the low $\mathcal{B}(H\to e^+e^-) = 5.22\times10^{-9}$ branching fraction and the large irreducible Drell--Yan continuum background, which result in a signal-to-background ratio of $\mathcal{O}(10^{-11})$, far beyond experimental reach.
The non-observation of $H \to e^+e^-$ at the LHC so far, sets an upper bound on the electron Yukawa coupling modifier, $\kappa_e=|y_e/y_e^{\rm \textsc{sm}}| \lesssim 260$~\cite{Khachatryan:2014aep,ATLAS:2019old}.
Projections indicate that the HL-LHC will improve this limit to $\kappa_e \lesssim 120$ for an integrated luminosity of 3\,ab$^{-1}$~\cite{Cepeda:2019klc}.
Achieving significantly stronger constraints on $\kappa_e$ is therefore a central goal of Higgs precision studies at future high-energy colliders~\cite{Altmann:2025feg,deBlas:2025gyz}.
In particular, the Future Circular Collider (FCC), both in its electron-positron (FCC-ee) and hadron-hadron (FCC-hh) operating phases~\cite{Abada:2019lih,Abada:2019zxq,FCC:2025lpp,FCC:2025uan,FCC:2025jtd}, offers a very rich Higgs physics program, allowing both stringent tests of the SM~\cite{Blondel:2018aan,Blondel:2019yqr,Azzi:2021gwg,Azzurri:2021nmy,DelVecchio:2025gzw,Kahraman:2025myp,Maura:2025rcv,FCCee_HiggsInvisible_2025,Kemp:2026mnh,Giappichini:2026vlg} 
and sensitive probes of new physics~\cite{Kamenik:2023hvi,Ripellino:2024iem,Cazzaniga:2025piw,Arroyo-Urena:2025mju,terHoeve:2025omu,Bhattacherjee:2025dlu,Elgammal:2026xdu}. 
In addition, it was early recognized that the FCC-ee provides the unique opportunity to probe the electron Yukawa coupling $y_e$ through a dedicated run at a center-of-mass (c.m.) energy of $\sqrt{s} = m_H$ exploiting $s$-channel Higgs production, $e^+e^- \to H$~\cite{dEnterria:2014,dEnterria:2017dac}. Despite a minuscule resonant production cross section~\cite{Jadach:2015cwa,Greco:2016izi}, and enormous irreducible backgrounds~\cite{dEnterria:2021xij}, several physics scenarios beyond the Standard Model (BSM) predict modifications of the electron Yukawa coupling~\cite{Altmannshofer:2015qra,Dery:2017axi,Ali:2021kxa,Solomon:2022qqf,Bahl:2022yrs,Chang:2022eft,Chang:2022pue,Davoudiasl:2023huk,Idegawa:2023bkh,Altmannshofer:2023tsa,Allwicher:2025mmc,Erdelyi:2025axy}  that could be only accessible through such a direct measurement.

A comprehensive generator-level analysis of resonant $s$-channel Higgs production at FCC-ee was performed in Ref.~\cite{dEnterria:2021xij} across eleven distinct decay final states, with the focus on quantifying the potential constraints on the electron Yukawa coupling, including accelerator considerations for its realization. The strategy of this study is based on a ``counting experiment'' approach that relies on selecting different final states in $\epem$ collisions at $\sqrts = m_H$ consistent with any of the Higgs decay modes, which produce a small ---yet potentially statistically significant when combined together--- excess in the measured cross sections over the background-only expectation. This study revealed that the digluon decay channel $H\to gg$ showed the best statistical significance, and the semileptonic $H\to WW^*\to \ell\nu+jj$ final state (where $\ell$ stands for charged leptons, and $jj$ for dijets) emerged as the second most promising signature. Including a $4.1$-MeV \cm\ energy spread, consistent with the SM Higgs boson total decay width of $\Gamma_{H} = 4.1$\,MeV~\cite{deFlorian:2016spz} and initial state radiation (ISR) effects, a signal significance of $1.3$\,standard-deviations (s.d.) could be achieved at 95\% confidence level (CL) by combining all decay channels and integrating 10~ab$^{-1}$ of data. This first generator-level investigation highlighted the challenging yet unique prospects for probing the electron Yukawa coupling at FCC-ee, was accompanied by parallel accelerator studies on beam monochromatization~\cite{Zimmermann:2017tjv,ValdiviaGarcia:2019ezi,Zhang:2024sao} towards reducing the beam energy spread (BES) to a minimum, by generating opposite correlations between spatial positions in the colliding beams, while preserving the highest possible luminosity. Simulations carried out in Ref.~\cite{Zhang:2024sao} investigated the optics required for a novel collision mode at the FCC-ee, presenting a study of monochromatization interaction region configurations and their implementation within the global FCC-ee lattice design for $s$-channel Higgs production. Aspects of the proposed monochromatization parameters have been discussed and critically examined in subsequent works~\cite{Shatilov:2025fck}.
Additional studies~\cite{Boughezal:2024yjk} showed, among others, that adding transverse (longitudinal) polarizations to the $e^\pm$ beams could further enhance the sensitivity to $y_e$ by a factor of three (five) relative to unpolarized measurements, because of the reduced production of backgrounds in the $b\bar{b}$ and $WW^*$ channels, 
although achieving such highly monochromatized and polarized beams remains unproven.

\begin{figure*}[htbp!]
	\centering
	\includegraphics[width=0.99\textwidth]{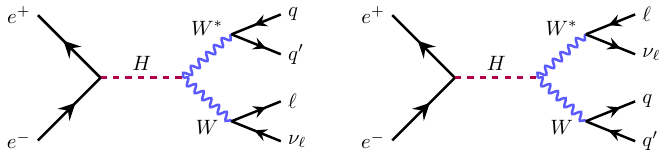}
		\caption{{\small Representative Feynman diagrams for the  $s$-channel Higgs boson production in $\epem$ collisions, with the Higgs decaying into a pair of $W$ bosons leading a lepton+jets final state. In the left (right) diagram, the on-shell (off-shell) $W$ decays leptonically to a charged lepton, $\ell=e,\mu,\tau$ and its corresponding neutrino, while the off-shell (on-shell) $W$ decays hadronically into two jets.} 
		\label{Fig:signal_diagram}}
\end{figure*}

Despite the formidable experimental challenges ---including the need for extremely narrow BES via monochromatization to match the narrow Higgs width, MeV-accurate knowledge of the Higgs boson mass position~\cite{Azzurri:2021nmy}, suppression of overwhelming backgrounds, and high integrated luminosity, as well as advanced detector requirements~\cite{Dam:2025zed}--- the potential for a direct, model-independent determination of $y_e$ provides strong motivation for a dedicated 125 GeV run at the FCC-ee. Such a measurement would close a crucial gap in our understanding of the Higgs mechanism and the origin of fermion masses. In this context, the present study aims to further improve and better quantify the possibility of probing the electron Yukawa coupling through resonant $s$-channel Higgs production in the semileptonic lepton+jets decay modes, $e^+e^-\to H \to WW^*\to\ell^\pm\nu+jj$ (Fig.~\ref{Fig:signal_diagram}) by using simulations based on the IDEA detector at the FCC-ee~\cite{IDEAStudyGroup:2025gbt,delphes_card_IDEA}. The analysis combines four separate signal channels, corresponding to electron and muon final states in both on- and off-shell $W$ boson configurations, with leptonic $\tau$ decays included in the corresponding $e,\,\mu$ channels. Figure~\ref{Fig:signal_diagram} illustrates the representative on- and off-shell diagrams for the final states of interest.
As default benchmark yields for the signal, we use $\sigma_{\epem\to H} = 280$\,ab~\cite{dEnterria:2021xij} as the resonant $e^+e^-\to H$ cross section (to be compared with the theoretical Breit--Wigner (BW) peak cross section of $1.64$\,fb~\cite{Jadach:2015cwa}) taking into account ISR effects and a monochromatized BES leading to a $\delta_{\sqrts} = 4.1$\,MeV \cm\ energy spread, and $\LumiInt = 10$\,ab$^{-1}$ for the achievable integrated luminosity~\cite{FCC:2025lpp}. A multiclass gradient boosted decision tree is employed to discriminate the signal from the dominant $WW^*$ continuum, $Z^*$, and $Z+X$ backgrounds, achieving a combined significance of $2.0$\,s.d.\ and an upper limit of $\kappa_e \lesssim 1.35$ at 95\% CL. The statistical significance corresponding to any other alternative running scenario $(\delta_{\sqrts},\LumiInt)$ can be derived by exploiting the bidimensional map derived in Ref.~\cite{dEnterria:2021xij}. The improved significance in these simulation studies with respect to the generator-level work of Ref.~\cite{dEnterria:2021xij} is driven by the use of four distinct signal categories, together with a more powerful machine-learning discriminant model.

The paper is organized as follows. Section~\ref{sec:Simulation} describes the Monte Carlo (MC) event generation and detector simulation. Analysis methodology is discussed in Sec.~\ref{sec:analysis}. The results and discussion are presented in Sec.~\ref{sec:Results}, and finally, Sec.~\ref{sec:Conclusion} summarizes our findings.

\section{Signal and background processes}
\label{sec:Simulation}

Matrix-element level events for signal and background processes in $\epem$ collisions at the Higgs pole are simulated with the \textsc{Whizard} MC event generator~\cite{Kilian:2007gr}. 
The data samples are produced within the official FCC software framework~\cite{francois_2025_kz4kp-h2j85}. Background events are generated including ISR photon emission, whereas signal events are generated without ISR 
but normalized to our benchmark cross section of $\sigma(\epem\to H)=280$~ab, accounting for the combined $\delta_{\sqrts} = 4.1$\,MeV BES and ISR effects~\cite{dEnterria:2021xij,Jadach:2015cwa}. 
To ensure statistically representative samples after the preselection and multivariate analysis (MVA) stages, meaningful numbers of events are generated for all processes, \eg up to $5 \times 10^8$ for the $Z^*\to jj$ process (with a cross section of $\mathcal{O}(10^3\, \mathrm{pb})$). Parton showering, hadronization, and particle decays are simulated using \textsc{Pythia~6}~\cite{Sjostrand:2006za}. Detector response is subsequently modeled with \textsc{Delphes}~\cite{deFavereau:2013fsa} using the IDEA detector configuration~\cite{IDEAStudyGroup:2025gbt}. The IDEA tracker provides coverage up to $|\cos\theta| < 0.99$, while the electromagnetic calorimeter extends to $|\cos\theta| < 0.995$. Electrons and muons are reconstructed within the tracker acceptance with 99\% efficiency if $E > 2\,\mathrm{GeV}$, while photon candidates are reconstructed within the ECAL acceptance with 99\% efficiency if $E > 2\,\mathrm{GeV}$. The complete simulation workflow is implemented within the \textsc{Key4HEP} framework~\cite{Key4hep:2023nmr}.

\begin{figure*}[htbp] 
	\centering
	\includegraphics[width=1.0\textwidth]{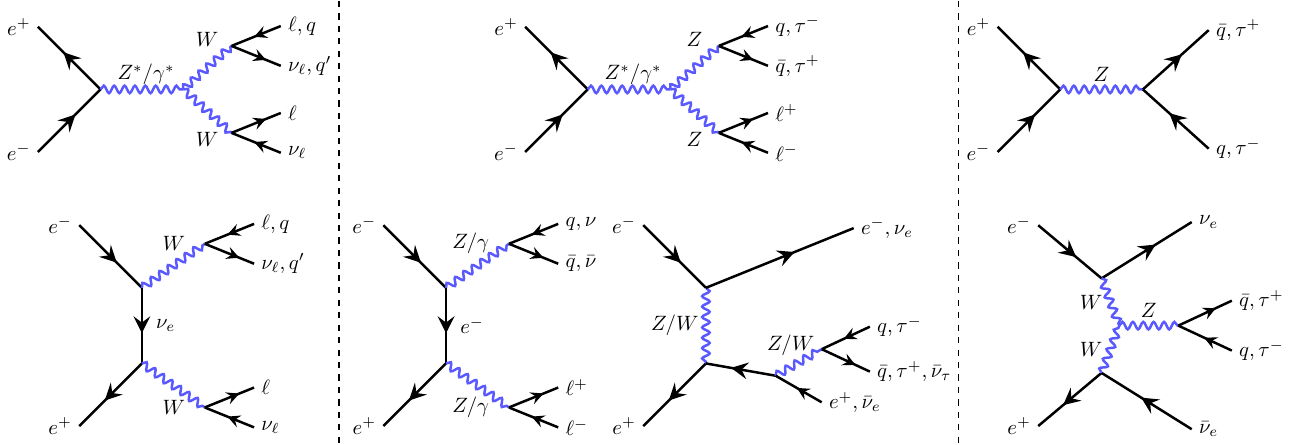}
    \caption{\small Representative Feynman diagrams for the dominant background processes in $\epem$ collisions at the Higgs pole: $WW^*$ continuum (left), $Z+X$ (middle), and $Z^*$ (right), producing leptons-plus-jets final states. 
\label{Fig:bkg_diagram}}
\end{figure*}

The dominant background processes for this study include $WW^*$, as well as $Z^*$ and $Z+X$ production (Fig.~\ref{Fig:bkg_diagram}). Among them, those with $e^\pm$ final states have larger cross sections than involving muons or tau leptons, because they include additional production channels with scattered $\epem$ from the colliding beams.
The $WW^*$ continuum background, particularly in the semileptonic final state ($\ell\nu jj$), closely mimics the signal kinematics and represents an irreducible background, with a production cross section of 
three (for $\mu\nu jj$) to four (for $e\nu jj$) orders-of-magnitude larger than the signal.
Fully leptonic $WW^* \to \ell\ell'\nu\nu$ (with $\ell = e, \mu, \tau$) decays are also included, as they can enter the selection through the hadronic decays of the $\tau$ lepton. The $Z+X$ background is simulated in the $2\ell 2j$, $2\ell 2\nu$, and $4\tau$ final states, with each lepton flavor generated separately. Events in $2\ell 2j$ contribute when jets mimic hadronic $W$ decays, $2\ell 2\nu$ provide missing energy overlapping the signal region, and $4\tau$ can enter via $\tau$-decays producing jets and leptons. In addition, the $Z^*\to jj$ and $Z^*\to \tau\tau$ processes are considered, whose production cross sections exceed that of the Higgs signal by $\mathcal{O}(10^{7})$ and $\mathcal{O}(10^{8})$, respectively. Although these processes do not share the same topology as the $H \to WW^*$ signal, their very large production rates make them relevant sources of reducible backgrounds. The $Z^*\to jj$ events can enter the selection through semileptonic heavy flavor decays, while $Z^*\to \tau\tau$ decays naturally produce charged leptons and neutrinos, leading to missing energy and partially overlapping with the signal signatures. 

In Table~\ref{tab:mc_samples}, the list of signal and background processes used in this study is presented (including their corresponding labels in the FCC \texttt{Winter2023} dataset for internal reference), as well as their production cross sections and number of generated MC events. The signal events are categorized according to on-shell and off-shell $W$ boson decays, with separate samples generated for each of the three lepton families, resulting in a total of six samples. For the $\tau$ signal samples, only leptonic $\tau$ decays are included.  
No theoretical uncertainties on the cross sections are considered, as they are expected to be subleading with respect to the statistical uncertainties of the measurements. After several years of FCC-ee operation at the $Z$ pole and $WW$ and $ZH$ \cm\ energies, the theoretical precision on the dominant background cross sections is expected to reach the $10^{-3}$ level or better~\cite{Blondel:2019vdq,Jadach:2018jjo,FCC:2025lpp}. Likewise, experimental systematic uncertainties (detector acceptance, reconstruction efficiencies, luminosity, etc.) are anticipated to be controlled at a similar level of precision, remaining well below statistical uncertainties~\cite{Abada:2019zxq,FCC:2025lpp}.

\begin{table}[htbp]
\centering
\caption{MC samples generated for the signal and background processes (with their corresponding labels in the FCC \texttt{Winter2023} dataset, for reference), associated cross sections (for the signal, a 280\,ab benchmark cross section normalization is assumed), and number of MC events generated.}
\label{tab:mc_samples}
\resizebox{\textwidth}{!}{%
\setlength{\tabcolsep}{22pt}
\begin{tabular}{llcc}
\hline\hline
Signal processes & label & cross section [ab] & MC sample size \\
\hline
 $H \to W(e\nu_e)W^*(jj)$& \texttt{Henueqq} & $4.38$ & $9.0 \times 10^{5}$ \\
 $H \to W(jj)W^*(e\nu_e)$ & \texttt{Hqqenue} & $4.38$ & $1.0 \times 10^{6}$ \\
 $H \to W(\mu\nu_\mu) W^*(jj)$ & \texttt{Hmunumuqq} & $4.38$ & $1.0 \times 10^{6}$ \\
 $H \to W(jj) W^*(\mu\nu_\mu)$ & \texttt{Hqqmunumu} & $4.38$ & $1.0 \times 10^{6}$ \\
 $H \to W(\tau\nu_\tau) W^*(jj)$& \texttt{Htaunutauqq} & $4.38$ & $1.0 \times 10^{6}$ \\ 
 $H \to W(jj) W^*(\tau\nu_\tau)$ & \texttt{Hqqtaunutau} & $4.38$ & $1.0 \times 10^{6}$ \\ \hline
 \multicolumn{4}{l}{Background processes} \\ \hline
$WW^* \to e\nu_e jj$& \texttt{enueqq} & $2.61 \times 10^{4}$ & $1.0 \times 10^{8}$ \\
$WW^* \to \mu\nu_\mu jj$& \texttt{munumuqq} & $6.71 \times 10^{3}$ & $1.0 \times 10^{8}$ \\
$WW^* \to \tau\nu_\tau jj$& \texttt{taunutauqq} & $6.76 \times 10^{3}$ & $1.0 \times 10^{8}$ \\
$WW^* \to \ell \ell' \nu\nu$& \texttt{l1l2nunu} & $9.85 \times 10^{3}$ & $1.0 \times 10^{8}$ \\ 
 $Z+X \to eejj$& \texttt{eeqq} & $3.93 \times 10^{6}$ & $1.0 \times 10^{8}$ \\
 $Z+X \to \mu\mu jj$& \texttt{mumuqq} & $1.51 \times 10^{5}$ & $1.0 \times 10^{8}$ \\
 $Z+X \to \tau\tau jj$& \texttt{tautauqq} & $1.48 \times 10^{5}$ & $1.0 \times 10^{8}$ \\
 $Z+X \to  4\tau$& \texttt{ZZ\_4tau} & $3.00 \times 10^{3}$ & $1.0 \times 10^{8}$ \\
 $Z+X \to ee\nu\nu$& \texttt{eenunu} & $6.57 \times 10^{5}$ & $1.0 \times 10^{8}$ \\
 $Z+X \to \mu\mu\nu\nu$& \texttt{mumununu} & $2.20 \times 10^{5}$ & $1.0 \times 10^{8}$ \\
 $Z+X \to \tau\tau\nu\nu$& \texttt{tautaununu} & $4.27 \times 10^{4}$ & $1.0 \times 10^{8}$ \\
 $Z^*\to \tau\tau$& \texttt{tautau} & $2.59 \times 10^{7}$ & $1.0 \times 10^{7}$ \\
$Z^*\to jj$ & \texttt{qq} & $3.63 \times 10^{8}$ & $5.0 \times 10^{8}$ \\
\hline\hline
\end{tabular}
}
\end{table}

\section{Analysis strategy}\label{sec:analysis}

The Higgs branching fraction to a pair of $W$ bosons is $\mathcal{B}(H\to WW^*) \approx 21.4\%$, the second largest after $H\to b\bar{b}$~\cite{Djouadi:2018xqq}. For semileptonic $WW^*$ decays, one $W$ boson decays hadronically with $\mathcal{B}(W\to jj) \approx 67.6\%$, while the other decays leptonically with $\mathcal{B}(W\to \ell\nu) \approx 32.4\%$, yielding an overall $\mathcal{B}(H\to WW^*\to \ell\nu jj) \approx 2\times 0.214\times 0.676\times 0.324 \simeq 9.4\%$, where the factor of two accounts for the two possible assignments of the leptonic and hadronic $W$ decays (charge conjugate modes)~\cite{ParticleDataGroup:2024cfk}. The analysis strategy exploits different categories based on the $W$ boson decay mode and whether they are on-shell or off-shell. In the first case, the on-shell $W$ decays into a charged lepton and its corresponding neutrino, whereas the off-shell one decays hadronically. In the second case, the off-shell $W$ decays leptonically, while the on-shell $W$ decays into two jets. 

Since various $H \to WW^* \to  \ell \nu jj$ subchannels exhibit different kinematic properties, each is treated individually. Four orthogonal event selections are therefore designed, corresponding to the electron and muon channels for both on- and off-shell configurations. The $W \to \tau\nu_\tau$ decays are also included in the signal definition, with only the leptonic $\tau$ decay channels taken into account. For simplicity in the rest of the manuscript, the symbol $\tau_\ell$ is used to denote those $\tau^-\to\ell^-\bar{\nu_\ell}\nu_\tau$ and $\tau^+\to\ell^+\bar{\nu}_\tau\nu_\ell$ (with $\ell=e,\mu$) decays. The same selection criteria are applied to the $\tau_\ell$ signal channels. These channels are then combined with the main electron and muon channels into four signal categories, as summarized below:
\begin{eqnarray*}
	\text{(I)}  &&  \qquad H\, \to\, W(e\nu_e)\, W^*(jj)\,\,\,+\,\,W(\tau_e\nu_\tau)\, W^*(jj),\\
	\text{(II)} &&  \qquad H\, \to\, W(\mu\nu_\mu)\, W^*(jj)\,\,+\,\,W(\tau_\mu\nu_\tau)\, W^*(jj), 	\\
	\text{(III)} && \qquad  H\, \to\, W^*(e\nu_e)\, W(jj)\,\,\,+\,\,W^*(\tau_e\nu_\tau)\, W(jj),\\
	\text{(IV)}	&& \qquad H\, \to\, W^*(\mu\nu_\mu)\, W(jj)\,\,+\,\,W^*(\tau_\mu\nu_\tau)\, W(jj).
\end{eqnarray*} 
Although the hadronic $\tau$-decay branching fraction, $\mathcal{B}(\tau\to \mathrm{hadrons})\approx64.8\%$, is about twice larger than the leptonic one, its impact on the final resonant Higgs search is expected to be less significant than the lepton+jets final state studied here due to the large contribution of hadronic backgrounds. The hadronic $\tau$-decay channels, however, could benefit from a dedicated analysis approach, which is left for a future study.

\subsection{Preselection and orthogonality requirements}

\begin{figure}[htpb!]
\centering
\includegraphics[width=0.49\columnwidth]{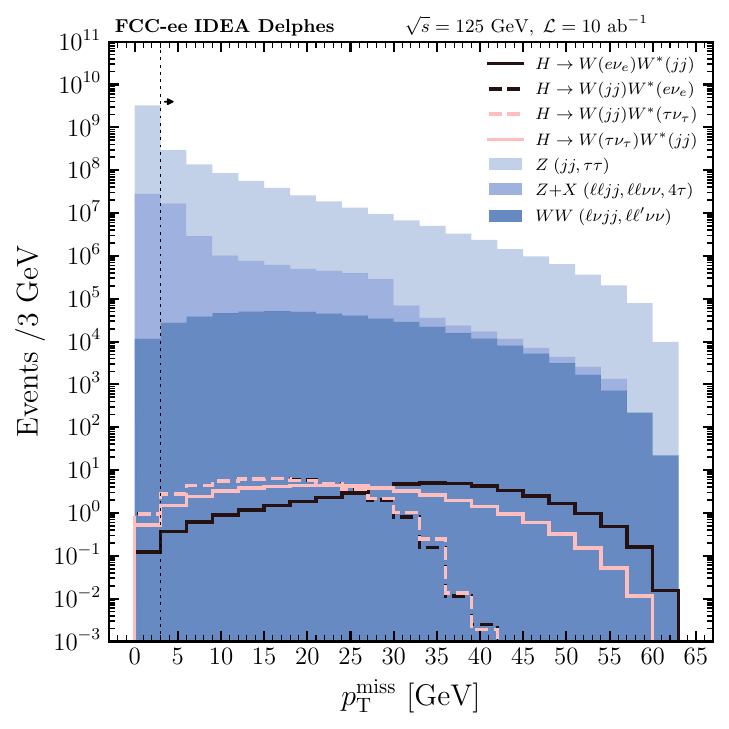}
\includegraphics[width=0.49\columnwidth]{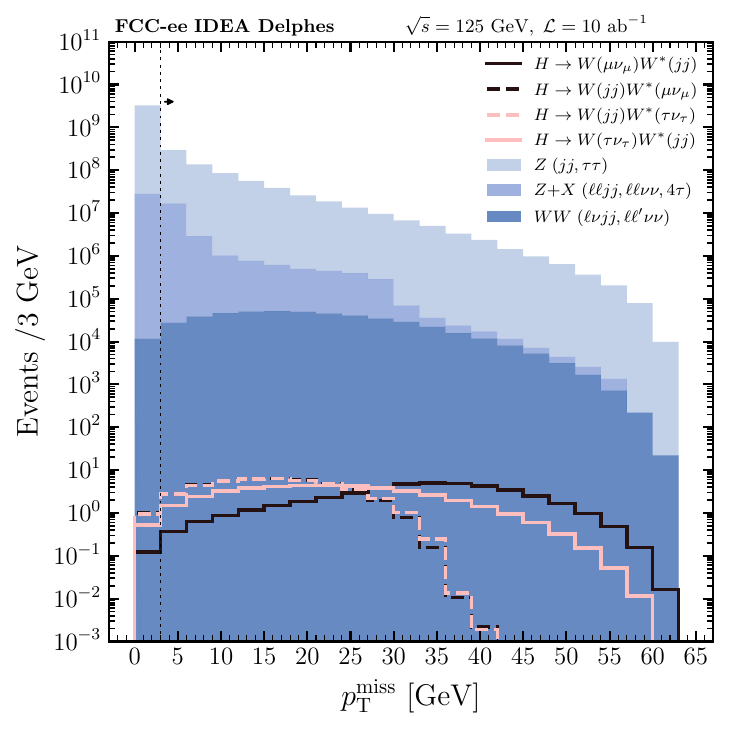}
\caption{\small Missing transverse momentum distributions for the electron (left) and muon (right) channels. Solid (dashed) black lines indicate the signal in the on-shell (off-shell) lepton channels. Pale pink solid (dashed) lines depict the signal in the on-shell (off-shell) $\tau_\ell$ channels. Filled stacked histograms show the backgrounds. Distributions are normalized to their cross sections for an integrated luminosity of 10\,ab$^{-1}$ at $\sqrt{s} = 125$\,GeV, with no selection cuts applied. The vertical dashed line indicates the $p_\mathrm{T}^{\rm miss}>3$\,GeV preselection cut.
\label{fig:met}}
\end{figure}

Three simple preselection kinematic criteria are defined and applied in the first stage of the analysis in order to remove simple reducible backgrounds while preserving the maximum amount of signal. The preselection procedure aims at reducing ``obvious'' background contributions while maximizing signal efficiency (above 90\%), thereby minimizing the number of background events used for the final MVA training and event selection. 
The first preselection criterion is based on a missing transverse momentum requirement of $p_\mathrm{T}^{\rm miss} > 3$\,GeV, due to the presence of a neutrino in the signal final-state of interest. The use of \textit{transverse}, rather than total, missing momentum is driven by the need to reduce the impact of background events with ISR activity, which can lead to a momentum imbalance along the longitudinal (but not transverse) direction. As seen in Fig.~\ref{fig:met}, this first cut suppresses the sharp $Z$-boson peaks of the $Z^*$ and $Z+X$ backgrounds. Radiative return events to the $Z$ boson, which are characterized by large missing longitudinal momentum $p_{z}^{\rm miss}$, but negligible missing transverse momentum $p_\mathrm{T}^{\rm miss}$, remain distinct from the signal. 
By applying this cut, nearly all signal events are retained, while more than 86\% of the $Z^*\to jj$ background, which has the largest cross section, is rejected. 
A substantial suppression of the $Z+X\to \ell\ell jj$ 
background is also obtained, as shown in Tables~\ref{table:presel_on_shell_electron}--\ref{table:presel_off_shell_muon}.

As a second preselection criterion, each event is required to have exactly one isolated charged lepton, \ie one isolated electron for the electron channels, and one isolated muon for the muon channels. To satisfy the isolation requirement, the relative cone momentum fraction $I_\mathrm{rel}=\sum_i \, p_\mathrm{T}^{\,i}/p_\mathrm{T}^{\,\ell}$, where $\ell$ is the candidate charged lepton and $i$ is the summation index running over charged hadrons (rather than over all particle-flow candidates to avoid removing genuinely isolated charged leptons with final state radiation (FSR) photons), is required to be below 0.2 within a cone of radius $R=0.2$ around the electron or muon. The choices of $I_\mathrm{rel}$ and $R$ values have been optimized to maximize signal selection efficiency while reducing the $Z^*$ and $Z+X$ backgrounds contributions, as explained in Appendix~\ref{App:A} in more detail. 

Figure~\ref{fig:isoCut} shows the distributions of the number of isolated electrons (muons) on the left (right) panel for signals (black lines) and backgrounds (colored histograms). Both plots exhibit a clear peak at $N_\ell^{\rm iso} = 1$ for the signal samples. 
Requiring events to contain only one isolated charged lepton (\ie\ within the vertical dashed lines) retains 95\% (92\%) of electron and muon on-shell (off-shell) signal channels, while removing a large fraction of the reducible $Z^*$ and $Z+X$ background contributions. In addition, approximately half (respectively, 10\%) of the $WW^*\to e\nu_e jj$ (respectively, $WW^*\to \mu\nu_\mu jj$) background events are rejected.

\begin{figure}[htpb!]
\centering
\includegraphics[width=0.49\columnwidth]{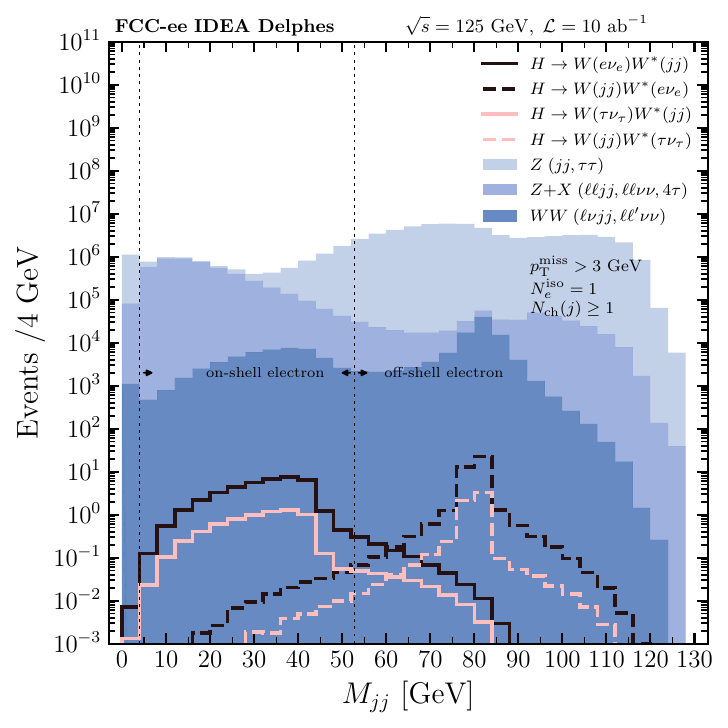}
\includegraphics[width=0.49\columnwidth]{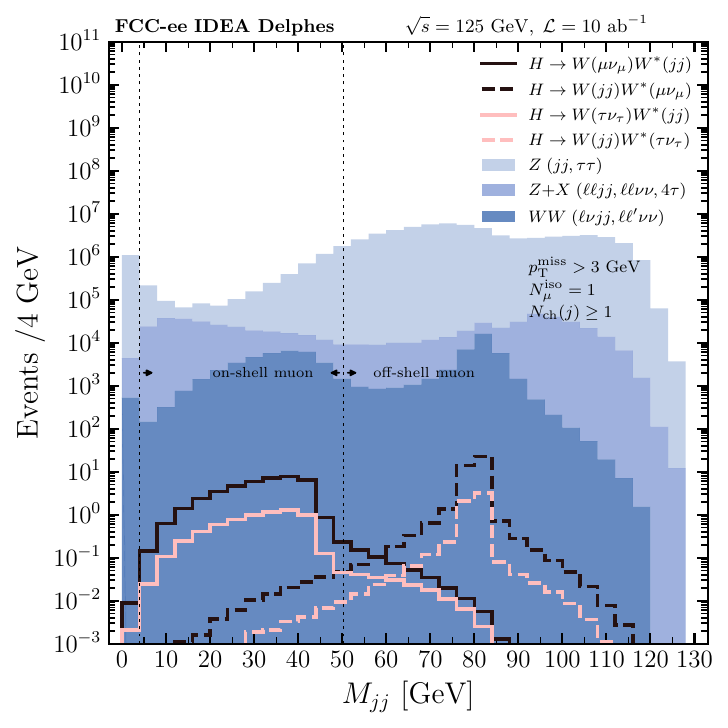}
\caption{\small Dijet invariant mass distributions for the electron (left) and muon (right) channels, after the first two preselection cuts ($p_\mathrm{T}^{\rm miss}>3$ and $N_\ell^{\rm iso} = 1$) applied. Jets are QCD jets, each of which contains at least 3 constituents (of which, at least, one charged hadron). Solid (dashed) black lines indicate the signal in the on-shell (off-shell) lepton channels. Pale pink solid (dashed) lines depict the signal in the on-shell (off-shell) $\tau_\ell$ channels. Filled stacked histograms show the backgrounds. Distributions are normalized to their cross sections for an integrated luminosity of 10\,ab$^{-1}$ at $\sqrt{s} = 125$\,GeV. The vertical dashed lines indicate the $4<M_{jj}<52.8$\,GeV (hadronic off-shell $W$) and  $M_{jj}>52.8$\,GeV (hadronic on-shell $W$) orthogonality preselection cuts.
\label{fig:IM}}
\end{figure}

Following the isolated lepton requirement, exactly two jets per event are reconstructed using the Durham-$k_\mathrm{T}$ clustering algorithm~\cite{Catani:1993hr} as implemented in the \textsc{FastJet} package~\cite{Cacciari:2011ma}. The algorithm is run on the full list of reconstructed particles in the event, excluding the isolated lepton candidate, and clusters all constituents into two exclusive jets ($N_j=2$).
Jets are required to have at least 3 constituents, one at least being a charged hadron ($N_\mathrm{ch}\geq 1$), in order to suppress backgrounds containing non-QCD jets and leptons, including taus. The third preselection criterion requires a minimum dijet invariant mass of $M_{jj} > 4\,\mathrm{GeV}$ in the event, to further reduce hadronic tau contributions.
To ensure orthogonality between the on- and off-shell signal categories, the invariant mass of the dijet system, $M_{jj}$, is used to separate the two channels. As seen in Fig.~\ref{fig:IM}, pairs of jets originating from an on-shell $W$ boson peak at $M_{jj} \approx 80$\,GeV, whereas jet pairs from an off-shell $W^*$ boson are typically at $M_{jj} \approx 40$\,GeV. Optimizing the separation threshold between the two categories results in values of $M_{jj} = 52.8$ and 50.3\,GeV for the dijet mass in the electron and muon channels, respectively. Events are then categorized as on-shell if $M_{jj}$ is below this mass threshold and off-shell otherwise, guaranteeing minimal overlap between the two channels and maximizing the discrimination power of the MVA stage. 

\begin{figure}[htpb!]
\centering
\includegraphics[width=0.49\columnwidth]{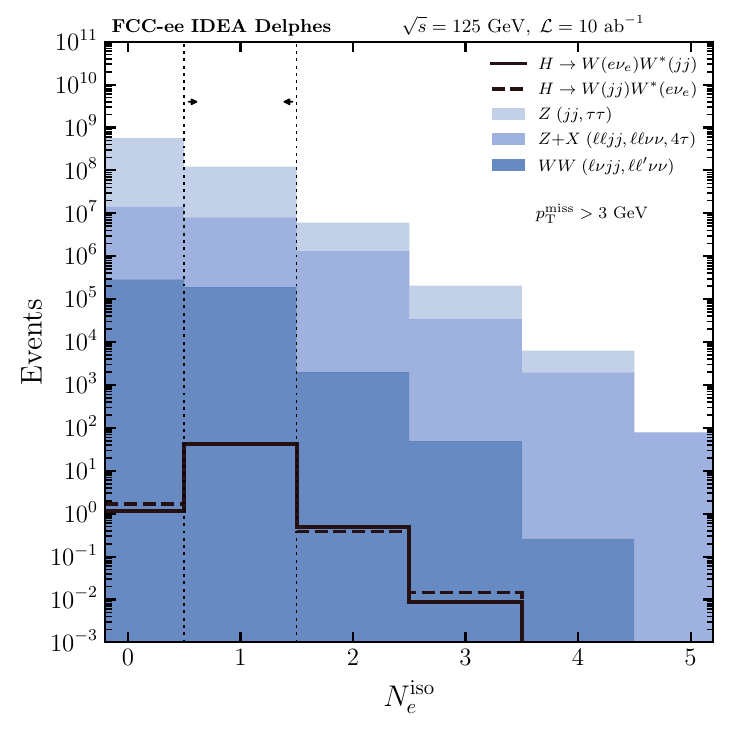}
\includegraphics[width=0.49\columnwidth]{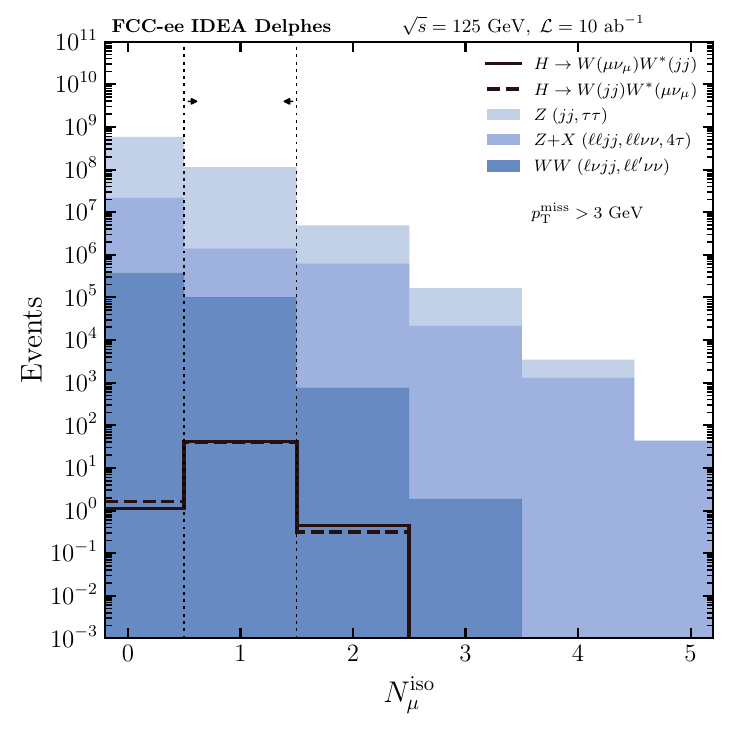}
\caption{\small Number of isolated charged leptons for the electron (left) and muon (right) channels, after the first preselection requirement ($p_\mathrm{T}^{\rm miss}>3$\,GeV) applied. Solid (dashed) black lines indicate the signal in the on-shell (off-shell) lepton channels. Filled stacked histograms show the backgrounds. Distributions are normalized to their cross sections for an integrated luminosity of 10\,ab$^{-1}$ at $\sqrt{s} = 125$\,GeV. The vertical dashed lines indicate the $N_\ell^{\rm iso} = 1$ preselection cut.
\label{fig:isoCut}}
\end{figure}

The same three-step preselection strategy is applied to the $\tau_\ell$ signal channels. Subsequently, for the MVA stage, events originating from on- and off-shell $\tau_{e}$ ($\tau_{\mu}$) decays into an electron (muon) are merged with the corresponding on- and off-shell electron (muon) signal samples. The event preselection results are summarized in the cut-flow Tables~\ref{table:presel_on_shell_electron}--\ref{table:presel_off_shell_muon} for the four categories of events, where the expected number of events and corresponding cumulative efficiency values for the signal and background processes are presented for the three preselection cuts.

As an example, in the electron on-shell case, after performing the preselection cuts, about 90\% of the signal events (as well as 87\% of the on-shell $\tau_e$ signal events) are retained, whereas the event selection efficiencies are around 16\% and 11\% for the $WW^* \to e\nu_e jj$ continuum and for the $Z+X \to eejj$ backgrounds, respectively. The overall preselection efficiency for $Z^*$ backgrounds is below 0.2\%. However, due to their large cross sections, the post-preselection yields of the reducible $Z^* \to jj$ and $Z+X \to eejj$ background processes remain still very important, amounting to $\mathcal{O}(10^6)$ events, compared to approximately 40 signal
events in each category. In general, the remaining backgrounds after preselection are larger for the off-shell than for the on-shell channels. This is primarily due to the $Z^*\to jj$ contamination, where heavy flavor decays produce soft leptons that more easily pass the selection in the off-shell categories, where the signal lepton is itself softer. Background yields are also smaller for the muon compared to the electron categories, due to the absence of additional
$t$-channel production diagrams involving the scattered beam leptons. To further enhance the discrimination between signal and background events, an MVA approach is employed next, as discussed in the following section.


\begin{table}[htbp!]
\centering
\caption{\small Number of events that pass each successive preselection cut (with cumulative percent efficiencies in parentheses) for the on-shell electron signal channel ($\tau_e$ stands for $\tau\to e\nu_e\nu_\tau$) and the main background processes, for $\epem$ collisions at $\sqrts=125$\,GeV with an integrated luminosity of $\mathcal{L}=10\,\mathrm{ab}^{-1}$.}
\vspace{0.2cm} 
\setlength{\tabcolsep}{18pt}
\resizebox{\textwidth}{!}{%

\begin{tabular}{l l l l}
\hline\hline
Process & $p_\mathrm{T}^{\text{miss}}>3$\,GeV & $N_e^{\text{iso}} = 1,N_\mu^{\text{iso}} = 0$ & $4<M_{jj}<52.8$\,GeV \\
\hline
$H \to W(e\nu_e) W^*(jj)$                & $44$ (99.7\%) & $42$ (95.0\%) & $39$ (89.9\%) \\
$H\to W(jj) W^*(e\nu_e)$                 & $43$ (97.8\%) & $40$ (92.2\%) & $0$ (0\%) \\
$H \to W(\tau_e\nu_\tau) W^*(jj)$        & $8$ (98.8\%)  & $7$ (89.7\%)  & $7$ (86.7\%) \\
$H \to W(jj) W^*(\tau_e\nu_\tau)$        & $8$ (97.8\%)  & $6$ (79.0\%)  & $0$ (0\%) \\ \hline
$WW^*\to e\nu_e jj$                      & $2.56\times10^{5}$ (98.2\%) & $1.36\times10^{5}$ (52.0\%) & $4.2\times10^{4}$ (16.2\%) \\
$WW^*\to \mu\nu_\mu jj$                  & $6.61\times10^{4}$ (98.5\%) & $19$ (0.03\%)               & $6$ (0.01\%) \\
$WW^*\to\tau\nu_\tau jj$                 & $6.58\times10^{4}$ (97.4\%) & $1.00\times10^{4}$ (14.8\%) & $4997$ (7.4\%) \\
$WW^*\to \ell\ell'\nu\nu$                & $9.27\times10^{4}$ (94.1\%) & $1.91\times10^{4}$ (19.4\%) & $27$ (0.03\%) \\
$Z+X\to eejj$                            & $1.55\times10^{7}$ (39.4\%) & $5.78\times10^{6}$ (14.7\%) & $4.44\times10^{6}$ (11.3\%) \\
$Z+X\to \mu\mu jj$                       & $1.81\times10^{5}$ (12.0\%) & $196$ (0.01\%)             & $47$ (0.003\%) \\
$Z+X\to \tau\tau jj$                     & $1.15\times10^{6}$ (77.8\%) & $2.32\times10^{5}$ (15.7\%) & $3.03\times10^{4}$ (2.1\%) \\
$Z+X\to ee\nu\nu$                        & $3.88\times10^{6}$ (59.0\%) & $1.02\times10^{6}$ (15.5\%) & $0$ (0\%) \\
$Z+X\to\mu\mu\nu\nu$                     & $1.61\times10^{6}$ (73.2\%) & $0$ (0\%)                   & $0$ (0\%) \\
$Z+X\to\tau\tau\nu\nu$                   & $3.43\times10^{5}$ (80.3\%) & $5.89\times10^{4}$ (13.8\%) & $141$ (0.03\%) \\
$Z+X\to 4\tau$                           & $2.69\times10^{4}$ (89.8\%) & $5.73\times10^{3}$ (19.1\%) & $2651$ (8.8\%) \\
$Z^*\to jj$                                & $4.61\times10^{8}$ (12.7\%) & $5.81\times10^{7}$ (1.6\%)  & $2.93\times10^{6}$ (0.08\%) \\
$Z^*\to\tau\tau$                           & $2.11\times10^{8}$ (81.6\%) & $4.84\times10^{7}$ (18.7\%) & $9.79\times10^{4}$ (0.04\%) \\
\hline\hline
\end{tabular}
}
\label{table:presel_on_shell_electron}
\end{table}

\begin{table}[htbp!]
\centering
\caption{\small Number of events expected to pass each successive preselection cut (with cumulative percent efficiencies in parentheses) for the off-shell electron signal channel ($\tau_e$ stands for $\tau\to e\nu_e\nu_\tau$) and the main background processes, for $\epem$ collisions at $\sqrts=125$\,GeV with an integrated luminosity of $\mathcal{L}=10\,\mathrm{ab}^{-1}$.}
\setlength{\tabcolsep}{18pt}
\resizebox{\textwidth}{!}{%

\begin{tabular}{l l l l }
\hline\hline
Process & $p_\mathrm{T}^{\text{miss}}>3$\,GeV & $N_e^{\text{iso}} = 1,N_\mu^{\text{iso}} = 0$ & $M_{jj}>52.8$\,GeV \\
\hline
$H \to W(e\nu_e) W^*(jj)$                & $44$ (99.7\%) & $42$ (95.0\%) & $1$ (1.8\%)   \\
$H\to W(jj) W^*(e\nu_e)$                 & $43$ (97.8\%) & $40$ (92.2\%) & $40$ (91.6\%)   \\
$H \to W(\tau_e\nu_\tau) W^*(jj)$        & $8$ (98.8\%)  & $7$ (89.7\%)  & $0$ (0\%)   \\
$H \to W(jj) W^*(\tau_e\nu_\tau)$        & $8$ (97.8\%)  & $6$ (79.0\%)  & $6$ (78.0\%)   \\ \hline
$WW^*\to e\nu_e jj$                      & $2.56\times10^{5}$ (98.2\%) & $1.36\times10^{5}$ (52.0\%) & $9.13\times10^{4}$ (34.9\%)  \\
$WW^*\to \mu\nu_\mu jj$                  & $6.61\times10^{4}$ (98.5\%) & $19$ (0.03\%) & $12$ (0.02\%)   \\
$WW^*\to\tau\nu_\tau jj$                 & $6.58\times10^{4}$ (97.4\%) & $1.00\times10^{4}$ (14.8\%) & $4879$ (7.2\%) \\
$WW^*\to \ell\ell'\nu\nu$                & $9.27\times10^{4}$ (94.1\%) & $1.91\times10^{4}$ (19.4\%) & $0$ (0\%)   \\
$Z+X\to eejj$                            & $1.55\times10^{7}$ (39.4\%) & $5.78\times10^{6}$ (14.7\%) & $8.94\times10^{4}$ (0.23\%)   \\
$Z+X\to \mu\mu jj$                       & $1.81\times10^{5}$ (12.0\%) & $196$ (0.01\%) & $146$ (0.01\%)   \\
$Z+X\to \tau\tau jj$                     & $1.15\times10^{6}$ (77.8\%) & $2.32\times10^{5}$ (15.7\%) & $1.93\times10^{5}$ (13.11\%)  \\
$Z+X\to ee\nu\nu$                        & $3.88\times10^{6}$ (59.0\%) & $1.02\times10^{6}$ (15.5\%) & $0$ (0\%)  \\
$Z+X\to\mu\mu\nu\nu$                     & $1.61\times10^{6}$ (73.2\%) & $0$ (0\%) & $0$ (0\%)  \\
$Z+X\to\tau\tau\nu\nu$                   & $3.43\times10^{5}$ (80.3\%) & $5.89\times10^{4}$ (13.8\%) & $0$ (0\%)  \\
$Z+X\to 4\tau$                           & $2.69\times10^{4}$ (89.8\%) & $5.73\times10^{3}$ (19.1\%) & $1581$ (5.3\%)   \\
$Z^*\to jj$                                & $4.61\times10^{8}$ (12.7\%) & $5.81\times10^{7}$ (1.6\%) & $5.36\times10^{7}$ (1.5\%)   \\
$Z^*\to\tau\tau$                           & $2.11\times10^{8}$ (81.6\%) & $4.84\times10^{7}$ (18.7\%) & $3.41\times10^{4}$ (0.01\%)  \\
\hline\hline
\end{tabular}
}
\label{table:presel_off_shell_electron}
\end{table}
\begin{table}[htbp!]
\centering
\caption{\small Number of events expected to pass each successive preselection cut (with cumulative percent efficiencies in parentheses) for the on-shell muon signal channel ($\tau_\mu$ stands for $\tau\to\mu\nu_\mu\nu_\tau$) and the main background processes, for $\epem$ collisions at $\sqrts=125$\,GeV with an integrated luminosity of $\mathcal{L}=10\,\mathrm{ab}^{-1}$.}
\setlength{\tabcolsep}{18pt}
\resizebox{\textwidth}{!}{%

\begin{tabular}{l l l l}
\hline\hline
Process & $p_\mathrm{T}^{\text{miss}}>3$\,GeV & $N_\mu^{\text{iso}} = 1, N_e^{\text{iso}} = 0$ & $4<M_{jj}<50.3$\,GeV   \\
\hline
$H \to W({\mu}\nu_{\mu}) W^*(jj)$          & $44$ (99.7\%) & $42$ (95.0\%) & $40$ (91.25\%) \\
$H\to W(jj) W^*({\mu}\nu_{\mu})$           & $43$ (97.8\%) & $40$ (92.4\%) & $0$ (0\%) \\
$H \to W(\tau_{\mu}\nu_\tau) W^*(jj)$      & $8$ (98.8\%)  & $7$ (90.2\%) & $7$ (85.1\%) \\
$H \to W(jj) W^*(\tau_{\mu}\nu_\tau)$      & $8$ (97.8\%)  & $6$ (78.0\%) & $0$ (0\%) \\\hline
$WW^*\to e\nu_e jj$                        & $2.56\times10^{5}$ (98.2\%) & $999$ (0.38\%) & $184$ (0.07\%) \\
$WW^*\to \mu\nu_\mu jj$                    & $6.61\times10^{4}$ (98.5\%) & $6.27\times10^{4}$ (93.4\%) & $2.9\times10^{4}$ (43.9\%)  \\
$WW^*\to\tau\nu_\tau jj$                   & $6.58\times10^{4}$ (97.4\%) & $9790$ (14.5\%) & $4827$ (7.1\%)   \\
$WW^*\to \ell\ell'\nu\nu$                  & $9.27\times10^{4}$ (94.1\%) & $2.82\times10^{4}$ (28.6\%) & $8$ (0.09\%) \\
$Z+X\to eejj$                              & $1.55\times10^{7}$ (39.4\%) & $1.10\times10^{5}$ (0.28\%) & $8.44\times10^{4}$ (0.21\%)  \\
$Z+X\to \mu\mu jj$                         & $1.81\times10^{5}$ (12.0\%) & $3.83\times10^{4}$ (2.5\%) & $1.20\times10^{4}$ (0.80\%)  \\
$Z+X\to \tau\tau jj$                       & $1.15\times10^{6}$ (77.8\%) & $2.28\times10^{5}$ (15.4\%) & $2.76\times10^{4}$ (1.9\%)  \\
$Z+X\to ee\nu\nu$                          & $3.88\times10^{6}$ (59.0\%) & $0$ (0\%) & $0$ (0\%)  \\
$Z+X\to\mu\mu\nu\nu$                       & $1.61\times10^{6}$ (73.2\%) & $3.53\times10^{5}$ (16.1\%) & $0$ (0\%) \\
$Z+X\to\tau\tau\nu\nu$                     & $3.43\times10^{5}$ (80.3\%) & $579$ (13.5\%) & $15$ (0.004\%) \\
$Z+X\to 4\tau$                             & $2.69\times10^{4}$ (89.8\%) & $5670$ (18.9\%) & $2463$ (8.2\%)  \\
$Z^*\to jj$                                  & $4.61\times10^{8}$ (12.7\%) & $5.66\times10^{7}$ (1.6\%) & $2.03\times10^{6}$ (0.06\%)  \\
$Z^*\to\tau\tau$                             & $2.11\times10^{8}$ (81.6\%) & $4.77\times10^{7}$ (18.4\%) & $6.07\times10^{4}$ (0.02\%)\\
\hline\hline
\end{tabular}
}
\label{table:presel_on_shell_muon}
\end{table}

\begin{table}[htbp!]
\caption{\small Number of events expected to pass each successive preselection cut (with cumulative percent efficiencies in parentheses) for the off-shell muon signal channel ($\tau_\mu$ stands for $\tau\to\mu\nu_\mu\nu_\tau$) and the main background processes, for $\epem$ collisions at $\sqrts=125$\,GeV with an integrated luminosity of $\mathcal{L}=10\,\mathrm{ab}^{-1}$.}
\setlength{\tabcolsep}{18pt}
\centering
\resizebox{\textwidth}{!}{%

\begin{tabular}{l l l l}
\hline\hline
Process & $p_\mathrm{T}^{\text{miss}}>3$\,GeV & $N_\mu^{\text{iso}} = 1, N_e^{\text{iso}} = 0$ & $M_{jj}>50.3$\,GeV   \\
\hline
$H \to W({\mu}\nu_{\mu}) W^*(jj)$          & $44$ (99.7\%) & $42$ (95.0\%) & $1$ (1.2\%) \\
$H\to W(jj) W^*({\mu}\nu_{\mu})$           & $43$ (97.8\%) & $40$ (92.4\%) & $40$ (91.8\%) \\
$H \to W(\tau_{\mu}\nu_\tau) W^*(jj)$      & $8$ (98.8\%)  & $7$ (90.2\%) & $0$ (0\%) \\ 
$H \to W(jj) W^*(\tau_{\mu}\nu_\tau)$      & $8$ (97.8\%)  & $6$ (78.0\%) & $6$ (77.0\%) \\\hline
$WW^*\to e\nu_e jj$                        & $2.56\times10^{5}$ (98.2\%) & $999$ (0.38\%) & $783$ (0.30\%) \\
$WW^*\to \mu\nu_\mu jj$                    & $6.61\times10^{4}$ (98.5\%) & $6.27\times10^{4}$ (93.4\%) & $3.23\times10^{4}$ (48.17\%) \\
$WW^*\to\tau\nu_\tau jj$                   & $6.58\times10^{4}$ (97.4\%) & $9790$ (14.5\%) & $4821$ (7.13\%) \\
$WW^*\to \ell\ell'\nu\nu$                  & $9.27\times10^{4}$ (94.1\%) & $2.82\times10^{4}$ (28.6\%) & $0$ (0\%) \\
$Z+X\to eejj$                              & $1.55\times10^{7}$ (39.4\%) & $1.10\times10^{5}$ (0.28\%) & $4881$ (0.01\%) \\
$Z+X\to \mu\mu jj$                         & $1.81\times10^{5}$ (12.0\%) & $3.83\times10^{4}$ (2.5\%) & $2.44\times10^{4}$ (1.6\%) \\
$Z+X\to \tau\tau jj$                       & $1.15\times10^{6}$ (77.8\%) & $2.28\times10^{5}$ (15.4\%) & $1.91\times10^{5}$ (13.0\%) \\
$Z+X\to ee\nu\nu$                          & $3.88\times10^{6}$ (59.0\%) & $0$ (0\%) & $0$ (0\%) \\
$Z+X\to\mu\mu\nu\nu$                       & $1.61\times10^{6}$ (73.2\%) & $3.53\times10^{5}$ (16.1\%) & $0$ (0\%) \\
$Z+X\to\tau\tau\nu\nu$                     & $3.43\times10^{5}$ (80.3\%) & $579$ (13.5\%) & $0$ (0\%) \\
$Z+X\to 4\tau$                             & $2.69\times10^{4}$ (89.8\%) & $5670$ (18.9\%) & $1704$ (5.7\%) \\
$Z^*\to jj$                                  & $4.61\times10^{8}$ (12.7\%) & $5.66\times10^{7}$ (1.6\%) & $5.42\times10^{7}$ (1.5\%)  \\
$Z^*\to\tau\tau$                             & $2.11\times10^{8}$ (81.6\%) & $4.77\times10^{7}$ (18.4\%) & $78$ ($3.01\times10^{-5}$\%) \\
\hline\hline
\end{tabular}
}
\label{table:presel_off_shell_muon}
\end{table}

\clearpage
\subsection{Multi-class GBDT analysis}\label{sec:bdt}

In this section, the MVA method and the set of input observables used to optimize the separation between signal and background events are described. For the classification procedure, a multiclass Gradient Boosted Decision Tree (GBDT) algorithm~\cite{Xia:2018cfz} is implemented using the \textsc{XGBoost} package~\cite{Chen:2016:XGBoost}, and applied separately for each of the four channels, corresponding to on- and off-shell electron and muon signals. As aforementioned, signal events from $\tau$ decay channels where the $\tau$ decays to an electron ($\tau_e$) or a muon ($\tau_\mu$), have been combined with the corresponding electron and muon channels.

The classifier utilizes 95 kinematic and topological input features including isolated final-state single objects, missing four-momentum quantities representing the neutrino kinematics, as well as variables of reconstructed leptonic and hadronic $W$ boson candidates. In addition, Matrix Element Likelihood Analysis (MELA) angular variables~\cite{CMS:2012vby}, using the standard definitions of the decay angles in the Higgs rest frame, and event shape parameters such as sphericity, asphericity, planarity, and aplanarity, as well as global-event variables such as $M_\mathrm{event}$ (total invariant mass of the event) and $E_\mathrm{miss}$, are included in the training.
A jet flavor-tagging algorithm~\cite{Bedeschi:2022rnj} is employed to assign to each reconstructed jet a probability of originating from a given ``parton'' flavor ($u$, $d$, $s$, $c$, $b$, $g$, $\tau$). The tagging scores of the two jets provide discriminating power, since the signal features anticorrelated quark-flavors whereas the reducible background tends to exhibit jets with the same parton flavor. Moreover, the Durham $k_\mathrm{t}$ algorithm provides the recombination scales $d_{23}$ and $d_{34}$, defined as $d_{ij} = 2\,\min(E_i^2, E_j^2)\,(1 - \cos\theta_{ij})$, where $E_i$ and $E_j$ denote the particle energies and $\theta_{ij}$ is the angle between them. The quantities $d_{23}$ and $d_{34}$ correspond to the clustering scales at which the event would transition from a 3-jet to a 2-jet configuration and from a 4-jet to a 3-jet configuration, respectively. Although only two jets are ultimately reconstructed in an exclusive manner, this information is still included as input to the GBDT to improve the discrimination of the hadronic component of the events. An extended list of the MVA input variables, including their indicative discrimination ranking (Table~\ref{tab:BDT_full_vars}) is provided in Appendix~\ref{App:B} for the on-shell electron channel.

As illustrative examples, Figs.~\ref{fig:BDT_input} and~\ref{fig:BDT_input_norm} show representative distributions of several discriminating variables for the on- and off-shell electron channels. The former are normalized according to their production cross sections, total preselection efficiencies, and $\mathcal{L}_\mathrm{int}=10$\,ab$^{-1}$, whereas the latter shown the same distributions but normalized to unit area to enhance shape comparisons.
Additional distributions of input variables can be found in Figs.~\ref{fig:MVA_inputs} and~\ref{fig:MVA_inputs_norm} of Appendix~\ref{App:B}. Below, we briefly outline a few features of these distributions that help discriminate between signal and background:
%
\begin{figure}[htpb!] 
  \centering  
    \begin{subfigure}[b]{0.41\textwidth}
    \centering
    \includegraphics[width=\linewidth]{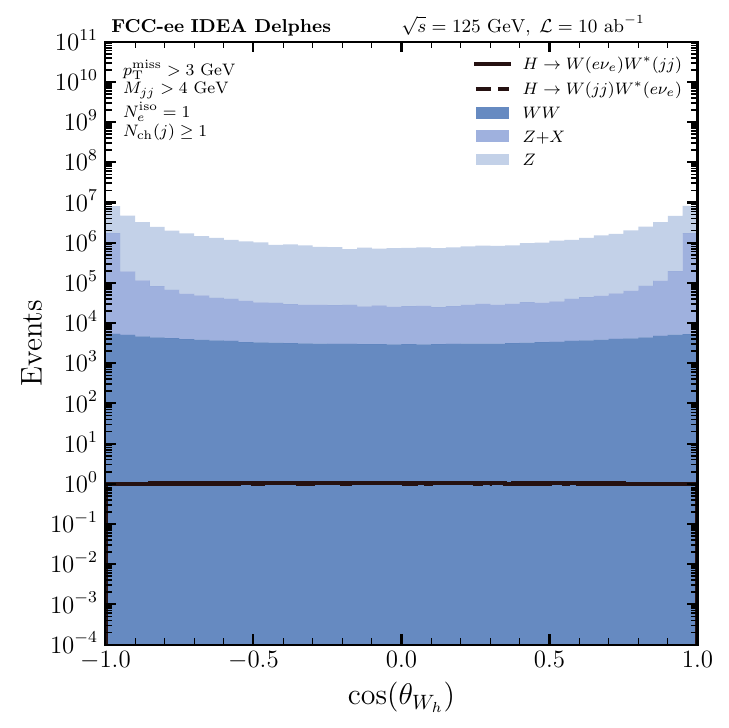}
    \captionsetup{skip=0pt}
    \caption{}
    \label{fig:cosw}
    \end{subfigure}
    \begin{subfigure}[b]{0.41\textwidth}
    \centering
\includegraphics[width=\textwidth]{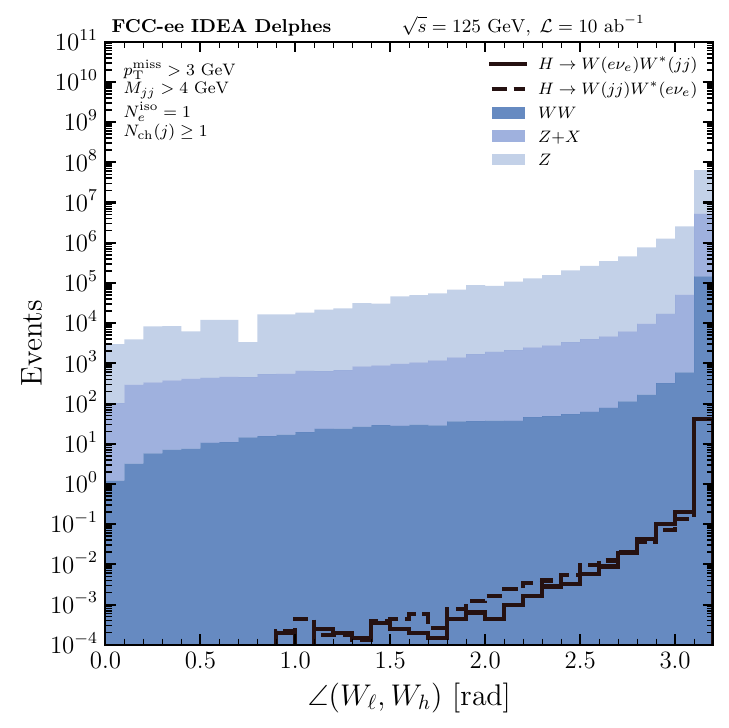}
    \captionsetup{skip=0pt}
    \caption{}
    \label{fig:WW_angle}
    \end{subfigure}
    \begin{subfigure}[b]{0.41\textwidth}
    \centering
    \includegraphics[width=\linewidth]{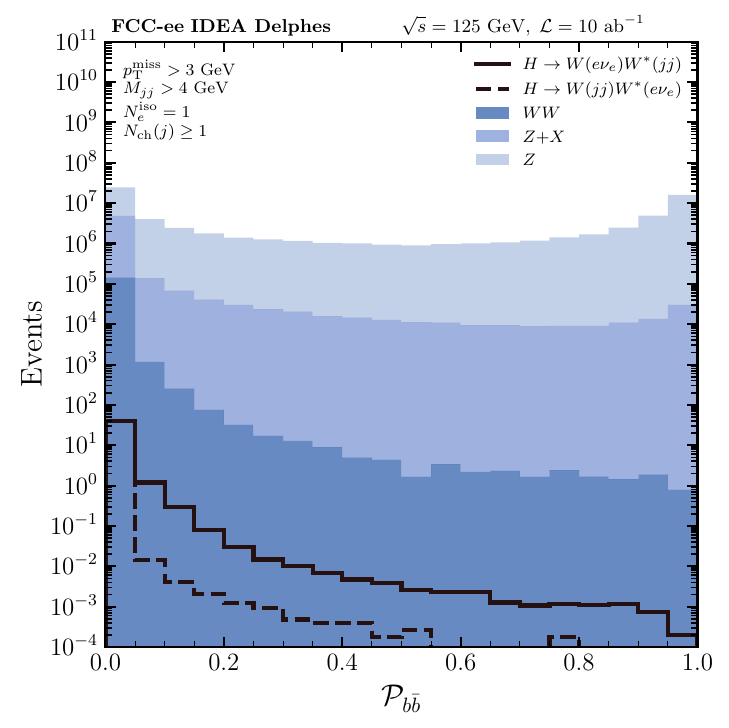}
    \captionsetup{skip=0pt}
    \caption{}
    \label{fig:bb}
    \end{subfigure}
    \begin{subfigure}[b]{0.41\textwidth}
    \centering
    \includegraphics[width=\linewidth]{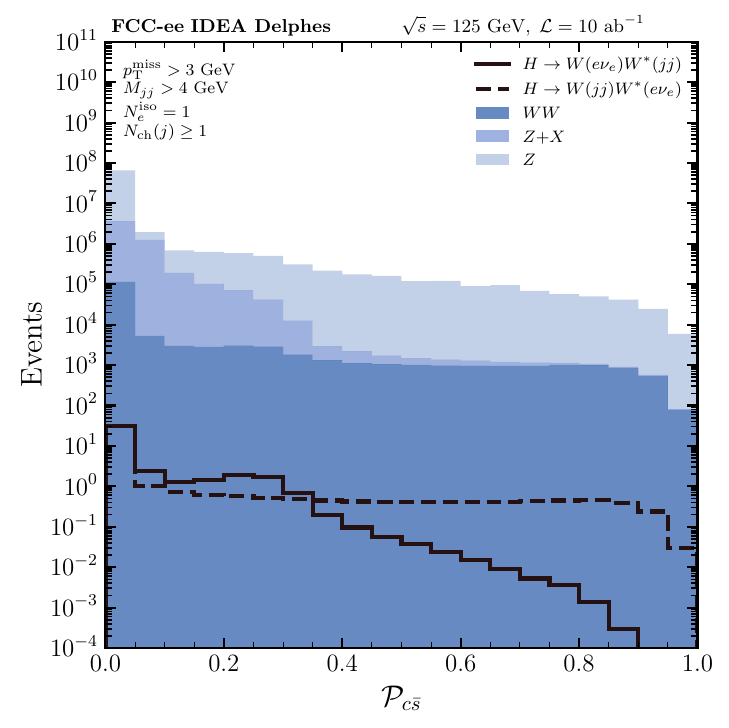}
    \captionsetup{skip=0pt}
    \caption{}
    \label{fig:cs}
    \end{subfigure}
    \begin{subfigure}[b]{0.41\textwidth}
    \centering
    \includegraphics[width=\linewidth]{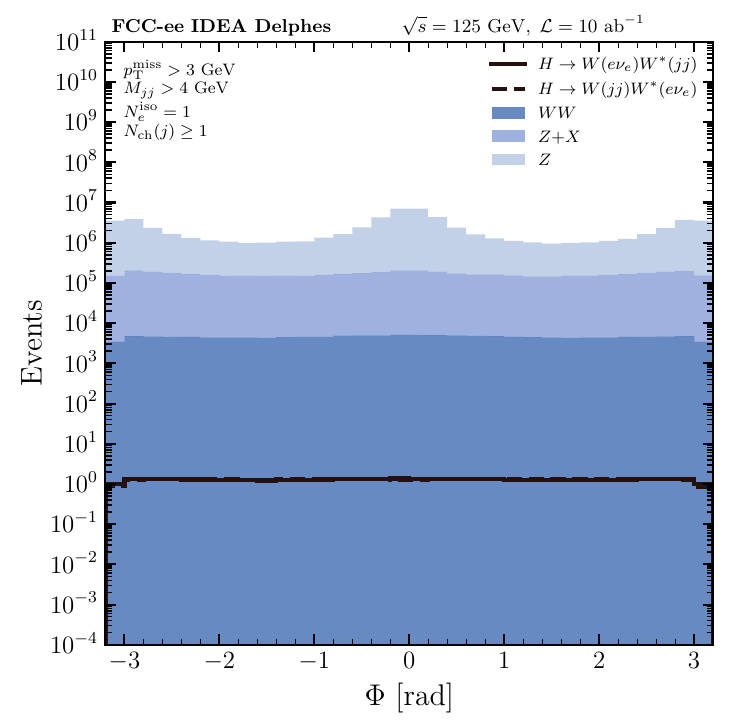}
    \captionsetup{skip=0pt}
    \caption{}
    \label{fig:MELA_Phi}
    \end{subfigure}
    \begin{subfigure}[b]{0.41\textwidth}
    \centering
    \includegraphics[width=\linewidth]{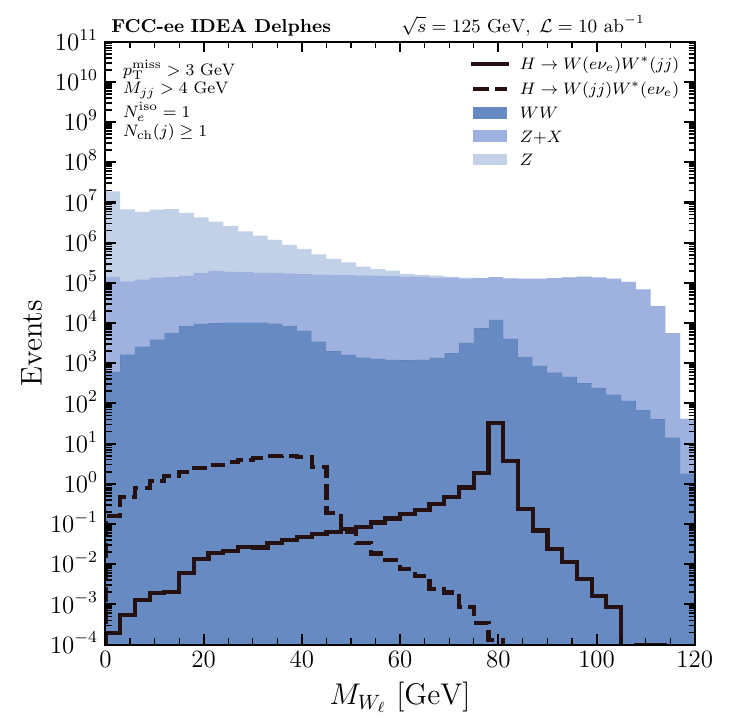}
    \captionsetup{skip=0pt}
    \caption{}
    \label{fig:W_lep_IM}
    \end{subfigure}
\caption{{\small Representative distributions of the MVA input variables for the on-shell (solid lines) and off-shell (dashed lines) electron signal channels versus background processes (filled stacked histograms) for $\epem$ collisions at $\sqrt{s}=125$\,GeV: (a) $\cos\theta_{W_h}$ of the reconstructed hadronic $W$ ($\theta_{W_h}$ defined with respect to the beam axis); (b) angle between the $W_\ell$ and $W_h$ bosons; (c) product of the $b$-tagging scores of the two jets, $\mathcal{P}_{b\bar{b}}$; (d) flavor-tagging product under the $c$-$s$ hypothesis, $\mathcal{P}_{cs}$; (e) $\Phi$ MELA angular variable; (f) invariant mass of the leptonically decaying $W$.}
\label{fig:BDT_input}}
\end{figure}


\begin{figure}[htpb!] 
  \centering  
    \begin{subfigure}[b]{0.41\textwidth}
    \centering
    \includegraphics[width=\linewidth]{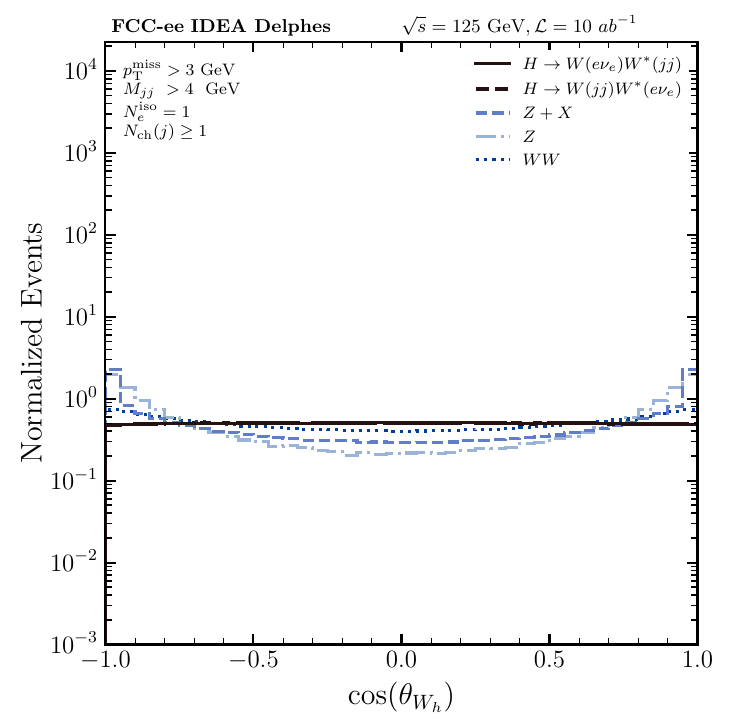}
    \captionsetup{skip=0pt}
    \caption{}
    \label{fig:cosw_norm}
    \end{subfigure}
    \begin{subfigure}[b]{0.41\textwidth}
    \centering
\includegraphics[width=\textwidth]{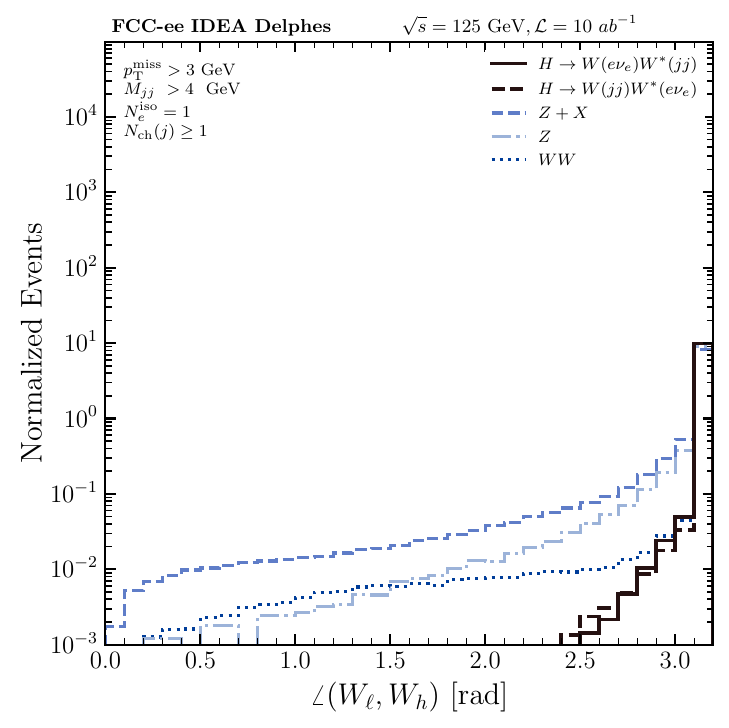}
    \captionsetup{skip=0pt}
    \caption{}
    \label{fig:WW_angle_norm}
    \end{subfigure}
    \begin{subfigure}[b]{0.41\textwidth}
    \centering
    \includegraphics[width=\linewidth]{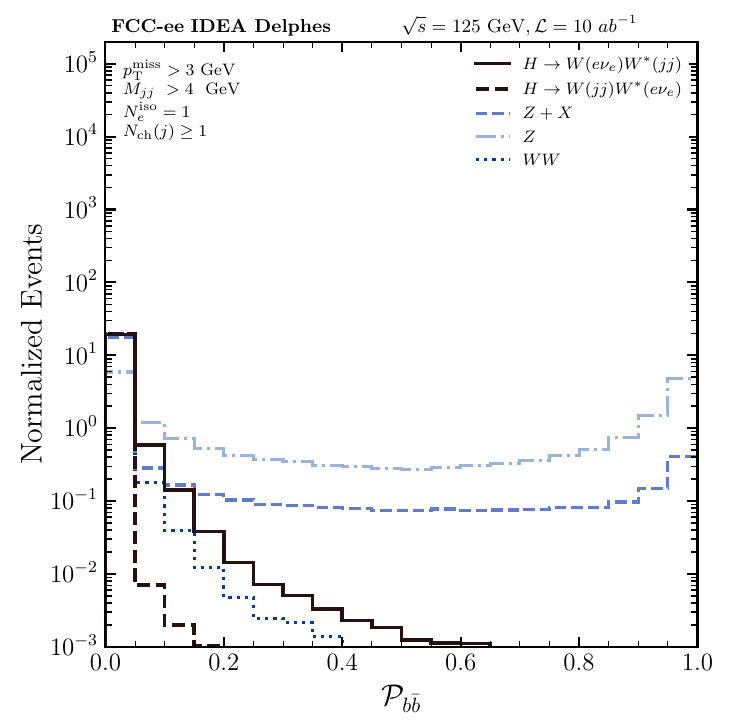}
    \captionsetup{skip=0pt}
    \caption{}
    \label{fig:bb_norm}
    \end{subfigure}
    \begin{subfigure}[b]{0.41\textwidth}
    \centering
    \includegraphics[width=\linewidth]{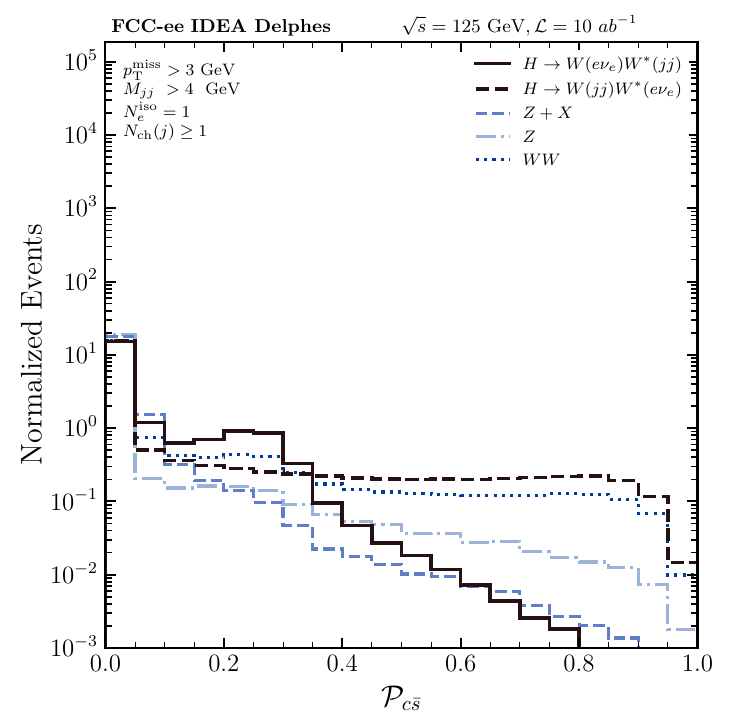}
    \captionsetup{skip=0pt}
    \caption{}
    \label{fig:cs_norm}
    \end{subfigure}
    \begin{subfigure}[b]{0.41\textwidth}
    \centering
    \includegraphics[width=\linewidth]{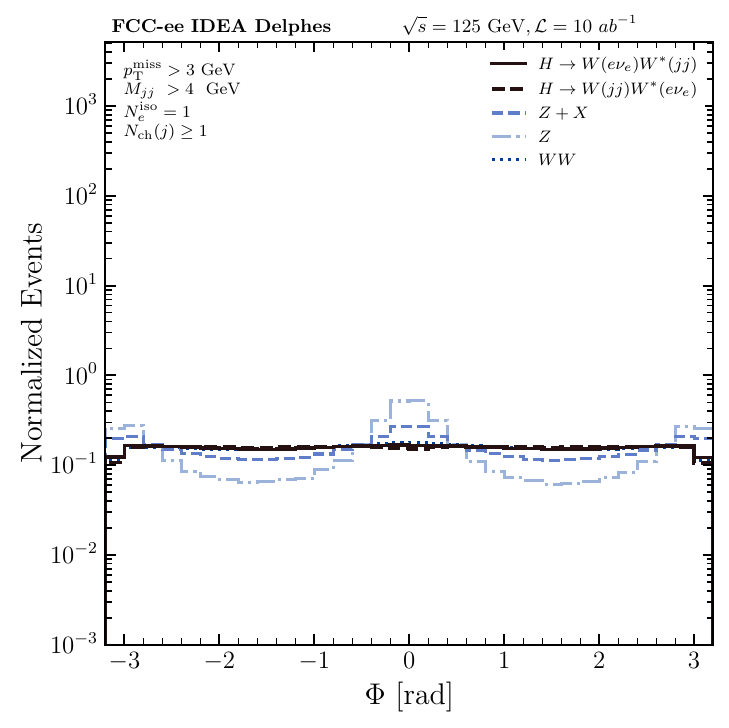}
    \captionsetup{skip=0pt}
    \caption{}
    \label{fig:MELA_Phi_norm}
    \end{subfigure}
    \begin{subfigure}[b]{0.41\textwidth}
    \centering
    \includegraphics[width=\linewidth]{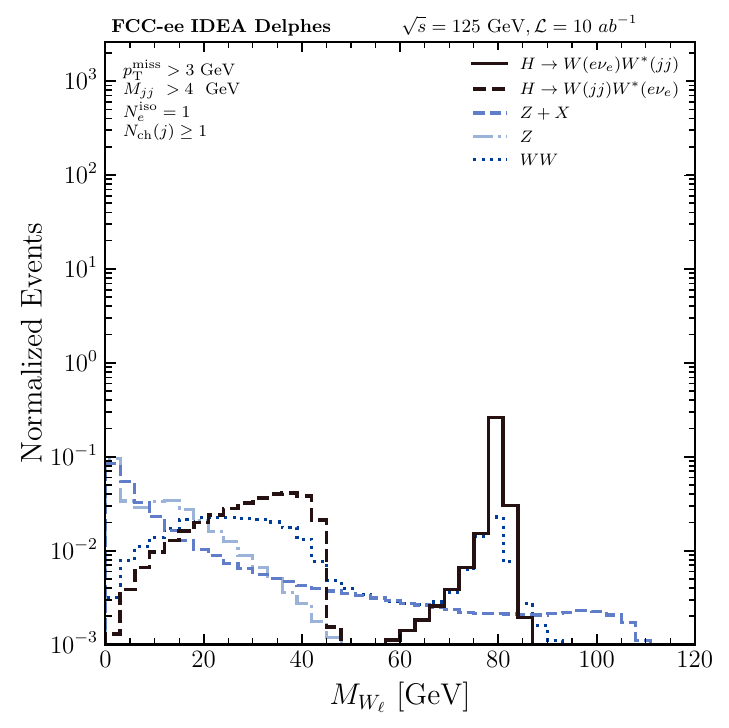}
    \captionsetup{skip=0pt}
    \caption{}
    \label{fig:W_lep_IM_norm}
    \end{subfigure}
\caption{\small 
Distributions of the same representative MVA input variables shown in Fig.~\ref{fig:BDT_input}, normalized to unit area to highlight differences in shape, for the on-shell (solid black histograms) and off-shell (dashed black histograms) electron signal channels and for the backgrounds (dotted, dashed, and dashed-dotted blue histograms).
\label{fig:BDT_input_norm}}
\end{figure}


%
\begin{itemize}[itemsep=0.1cm,parsep=0.1cm,topsep=0.1cm]
\item Figures~\ref{fig:cosw} and~\ref{fig:cosw_norm}: The $\cos\theta$ distribution (with $\theta$ defined with respect to the beam axis) of the dijet system, corresponding to the hadronically decaying reconstructed $W$ boson, exhibits a flat behaviour for the signal, consistent with the decay of a spin-zero Higgs boson. The $WW^*$ background (proceeding through spin-1 $s$-channel production, or through $t$-channel processes, see Fig.~\ref{Fig:bkg_diagram}) shows an increasingly concave shape for this observable, whereas the $Z^*$ and $Z+X$ processes are strongly enhanced in the forward-backward regions.

\item Figures~\ref{fig:WW_angle} and~\ref{fig:WW_angle_norm}: The angle between the two reconstructed leptonic and hadronic $W$ boson candidates, $\angle(W_\ell,W_h)$ is strongly peaked at $\pi$ for the signal, whereas it has a flatter tail towards to zero for all backgrounds. The signal
distribution is more sharply peaked because the Higgs boson is produced at rest, whereas background events allow for large (longitudinal) momentum ISR photons to be emitted, that impose a longitudinal
boost to the system and ultimately reduce the opening angle. 

%
%

\item Figures~\ref{fig:bb}, \ref{fig:cs} and~\ref{fig:bb_norm}, \ref{fig:cs_norm}: The jet-flavor tagging variables defined in this study are highly effective in separating signal from background, mainly due to the anticorrelated nature of jet flavors from $W$ boson hadronic decays, and the fact that the latter rarely produce (Cabibbo-suppressed) $b$ quarks, while background events typically contain two same-heavy-flavor jets, resulting in consistently high $b$-tagging scores for both jets. As a result, the product of the $b$-flavor tagging scores of the two jets, \ie the product probabilities of the two jets to be tagged as $b$-jets, $\mathcal{P}_{\bbbar}=\mathcal{P}_b(j_1)\cdot \mathcal{P}_b(j_2)$, shows a peak near unity for the reducible backgrounds (Figs.~\ref{fig:bb}  and~\ref{fig:bb_norm}), whereas the signals are suppressed significantly in this region, making this variable one of the most efficient discriminants in the analysis.
The distribution of the product of the flavor-tagging scores under the hypothesis that the first jet is charm-flavored and the second jet is strange-flavored, $\mathcal{P}_{c\bar{s}}=\mathcal{P}_c(j_1)\cdot \mathcal{P}_s(j_2)$ is shown in Figs.~\ref{fig:cs} and~\ref{fig:cs_norm}. The signal is enhanced at larger values of this observable when the on-shell $W$ decays hadronically (dashed line), especially compared to the $Z+X$ background, reflecting the Cabibbo-enhanced $W \to c\bar{s}$ decay\footnote{Of course, $W \to u\bar{d}$ is also a dominant hadronic decay mode, but the jet flavor tagging algorithms do not efficiently separate light quarks, unlike their performance for strange and heavy-flavor quarks~\cite{Bedeschi:2022rnj}.}. 

\item Figures~\ref{fig:MELA_Phi} and~\ref{fig:MELA_Phi_norm} show $\Phi$ MELA variable, defined as the dihedral angle between the two $W$ boson decay planes, with an orientation (sign) fixed by a chosen reference direction given by the isolated lepton. This variable is sensitive to both the spin correlations between the decay planes and the charge-conjugation and parity (CP) structure of the process (with the sign of $\Phi$ providing sensitivity to CP-odd effects). The $\Phi$ MELA distribution for the signal is found to be approximately flat except for distortions near the endpoints around $\pm\pi$, as expected for a spin-0 CP-even Higgs boson decay, whereas the $Z$-related backgrounds exhibit a clearly different behaviour, with pronounced enhancements around 0 and $\pm\pi$. The $WW^*$ continuum also features (small) modulations driven by different spin correlations than the signal.

%
\item Figures~\ref{fig:W_lep_IM} and~\ref{fig:W_lep_IM_norm} show the invariant mass distribution of the leptonic $W$ boson, $M_{W_\ell}$, which features peaks fully anticorrelated with those observed in the corresponding dijet invariant mass distribution (Fig.~\ref{fig:IM}), $M_{jj}\approx80$~(40)\,GeV, for the on-shell (off-shell) electron signal channel. 
Background contributions exhibit distinct behaviors across the leptonic invariant mass spectrum, with $Z+X$ events showing a distribution largely separated from the signal, with enhancements in the low-mass region ($<20$\,GeV) and the high-mass tail (80--110\,GeV). The $Z^*$ background is concentrated at very low $M_{W_\ell}$ values, as the lepton and missing momentum in these events arise from heavy flavor decays within jets.
\end{itemize}

For the GBDT training, four classes of events are considered: the resonant Higgs signal, the $WW^*$ continuum, and the $Z+X$ and $Z^*$ backgrounds. Feature importance is evaluated using gain-based metrics that measure the average improvement in purity contributed by each feature across all splits in the ensemble. The trained model achieves strong classification performance across all four leptonic categories. The \textsc{XGBoost} multiclass classification configuration, including the network parameters and class definitions, is described in Table~\ref{tab:network_config} of Appendix~\ref{App:B}.

\begin{figure}[htpb!]
\centering
\includegraphics[width=0.49\columnwidth]{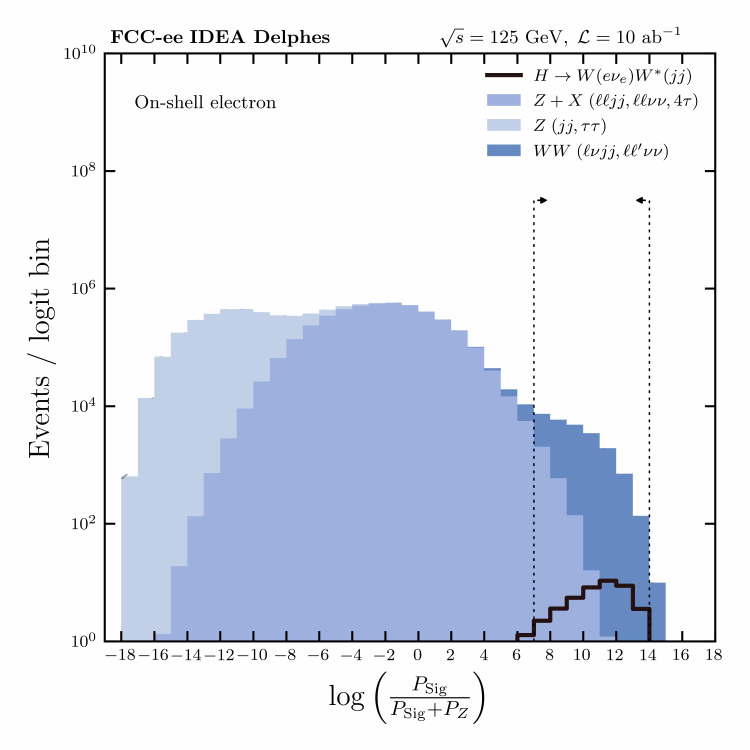}
\includegraphics[width=0.49\columnwidth]{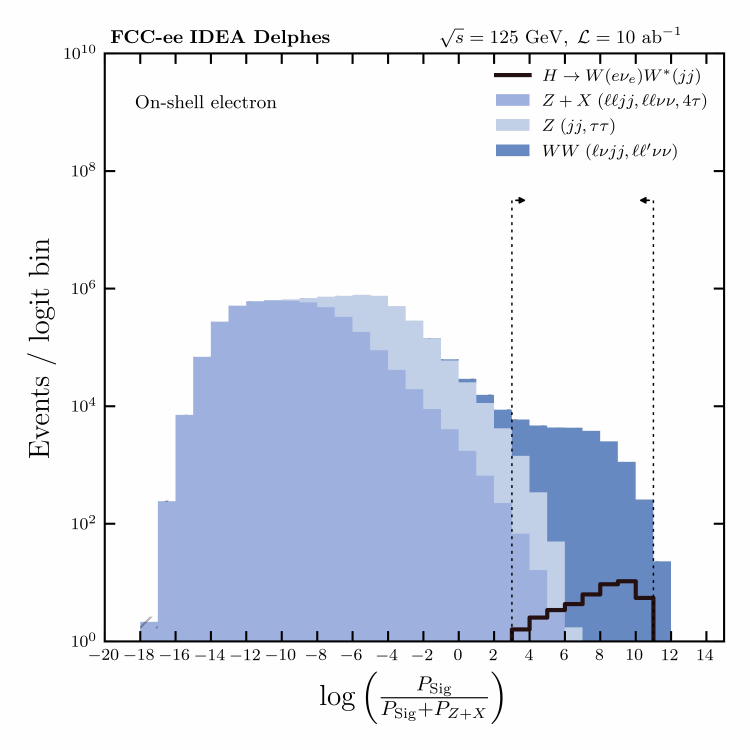}
\caption{\small 
{Distributions of the GBDT binary discriminants for the on-shell electron signal (solid black line) against the $Z^*$ (left) and $Z+X$ (right) backgrounds (filled histograms) events. Optimized signal-enhancing cuts are subsequently applied in the $[7,14]$ and $[4,11]$ ranges (vertical dashed lines) for the $Z^*$ and $Z+X$ backgrounds, respectively.} 
\label{fig:binary}}
\end{figure}

For each background, binary discriminants are defined as the probability of the signal divided by the sum of the probabilities for the signal and each background category, \ie\ $D_b = \mathcal{P}(\mathrm {sig})/(\mathcal{P}(\mathrm {sig})+\mathcal{P}(\mathrm {bkg}))$. The resulting distributions of $\log D_b$ for the on-shell electron signal with respect to the $Z+X$, and the $Z^*$ backgrounds are shown in Fig.~\ref{fig:binary}, as an example, illustrating their separation power between signal and reducible background events. Distributions for the remaining signal channels (off-shell electron, and on- and off-shell muon) are presented in Fig.~\ref{fig:other_binary} of Appendix~\ref{App:B}.
Following the training, the optimized GBDT model is applied to the full dataset in inference mode to generate classification scores for all simulated events.

Directly extracting the signal through a simultaneous fit in three dimensions is not straightforward and, therefore, optimized selections are applied first on the $Z^*$ and $Z+X$ binary discriminants, which efficiently suppress reducible backgrounds while preserving high signal efficiency. As shown in Fig.~\ref{fig:binary}, to maximize the sensitivity of the signal with respect to the backgrounds, events are selected within the intervals $[7,14]$ and $[4,11]$ of the binary discriminants, which remove the $Z^*$ and $Z+X$ backgrounds, respectively. The yields obtained after these selections for all four signal channels are summarized in Table~\ref{tab:yields}.
As observed, the majority of the reducible backgrounds are removed by applying the MVA selections. For the on-shell channels, all $Z^*$ background events are effectively eliminated, whereas in the off-shell channels, a small fraction of $Z^*\to jj$ backgrounds remains; though this residual contribution is negligible compared to the $WW^*$ continuum backgrounds (darker blue histograms in Figs.~\ref{fig:binary} and~\ref{fig:other_binary}). Similarly, a small number of $Z+X\to\ell\ell jj$ background events survive this first stage of the GBDT analysis. At the same time, most of the signal events are retained in all four categories. The on-shell lepton channels exhibit slightly better discrimination than the off-shell channels, as expected. 

From the number of remaining signal and background events presented in Table~\ref{tab:yields}, one can see that the irreducible $WW^*$ continuum background events left at this stage amounts to 73\,000 and 44\,000 for the $WW^*\to e\nu_e jj$ and  $WW^*\to \mu\nu_\mu jj$ categories (including both on- and off-shell contributions), respectively. These should be compared to the initial yields of 260\,000 and 67\,000 events, obtained by multiplying the generator-level cross section values of Table~\ref{tab:mc_samples} by 10\,ab$^{-1}$. Namely, the preselection and GBDT-based reducible-background analysis also remove a fraction of the continuum backgrounds, by factors of 0.28 and 0.65 for the electron and muon final states, respectively. The bottom row of Table~\ref{tab:mc_samples} lists the approximate statistical significance for each category, defined as the number of signal events divided by the square root of the sum of backgrounds remaining after this first step of the GBDT analysis. At this stage, the on-shell channels show a combined significance of about 0.4\,s.d., whereas the off-shell ones yield about 0.3\,s.d.\ (with the off-shell electron being affected by larger background contributions). As a final step, a likelihood fit is performed to extract the signal contribution over the remaining backgrounds, mostly $WW^*$ continuum, as discussed in the following section.


\begin{table}[htbp!]
\caption{\small Number of signal and background events per channel after applying the selection cuts on the GBDT binary discriminants to reduce the $Z^*$ and $Z+X$ backgrounds, for $\epem$ collisions at $\sqrts=125$\,GeV with an integrated luminosity of $\mathcal{L}=10\,\mathrm{ab}^{-1}$. The bottom row shows the approximate statistical significance, defined as the number of signal events divided by the square-root of the sum of remaining background events ($S_H/\sqrt{\sum_i B_i}$), 
for each category.
\label{tab:yields}}
\centering
\setlength{\tabcolsep}{10pt}
\resizebox{\textwidth}{!}{%

\begin{tabular}{lcccc}
\hline\hline
Process & on-shell e & off-shell e & on-shell $\mu$ & off-shell $\mu$ \\
\hline

\begin{tabular}{@{}l@{}}
$H \to W(e\nu_e) W^*(jj)$ \\
$H \to W(\tau_e\nu_\tau) W^*(jj)$\\
\hline
\end{tabular}
 & 41 & 0 & 0 & 0 \\
\begin{tabular}{@{}l@{}}
$H \to W(jj) W^*(e\nu_e)$ \\
$H \to W(jj) W^*(\tau_e\nu_\tau)$\\
\hline
\end{tabular}
 & 0 & 41 & 0 & 0 \\
\begin{tabular}{@{}l@{}}
$H \to W(\mu\nu_\mu) W^*(jj)$ \\
$H \to W(\tau_\mu\nu_\tau) W^*(jj)$\\
\hline
\end{tabular}
 & 0 & 0 & 43 & 0 \\
\begin{tabular}{@{}l@{}}
$H \to W(jj) W^*(\mu\nu_\mu)$ \\
$H \to W(jj) W^*(\tau_\mu\nu_\tau)$
\end{tabular}
 & 0 & 0 & 0 & 40 \\\hline
$WW^* \to e\nu_e jj$   & 19\,200 & 54\,000 & 1 & 20 \\
$WW^* \to \mu\nu_\mu jj$ & 0 & 2 & 23\,000 & 21\,200 \\
$WW^* \to \tau\nu_\tau jj$ & 840 & 1\,930 & 1\,590 & 1\,370 \\
$WW^* \to \ell \ell' \nu\nu$ & 0 & 0 & 0 & 0 \\ 
$Z+X \to eejj$ & 30 & 29 & 0 & 0 \\
$Z+X \to \mu\mu jj$ & 0 & 0 & 1 & 16 \\
$Z+X \to \tau\tau jj$ & 19 & 157 & 48 & 107 \\
$Z+X \to ee\nu\nu$ & 0 & 0 & 0 & 0 \\
$Z+X \to \mu\mu\nu\nu$ & 0 & 0 & 0 & 0 \\
$Z+X \to \tau\tau\nu\nu$ & 0 & 0 & 0 & 0 \\ 
$Z+X \to 4\tau$ & 0 & 0 & 0 & 0 \\
$Z^*\to \tau\tau$ & 0 & 0 & 0 & 0 \\
$Z^*\to jj$ & 0 & 626 & 0 & 60 \\ \hline
$S_H/\sqrt{\sum_i B_i}$ & $41/\sqrt{20\,090} \approx 0.29$ & $41/\sqrt{56\,740} \approx 0.17$ & $43/\sqrt{24\,640} \approx 0.27$ & $40/\sqrt{22\,770} \approx 0.26$ \\ 
\hline\hline
\end{tabular}
}
\end{table}

\clearpage
\section{Final result and discussion}
\label{sec:Results}

\begin{figure}[htpb!]
\centering
    \includegraphics[width=0.44\textwidth]{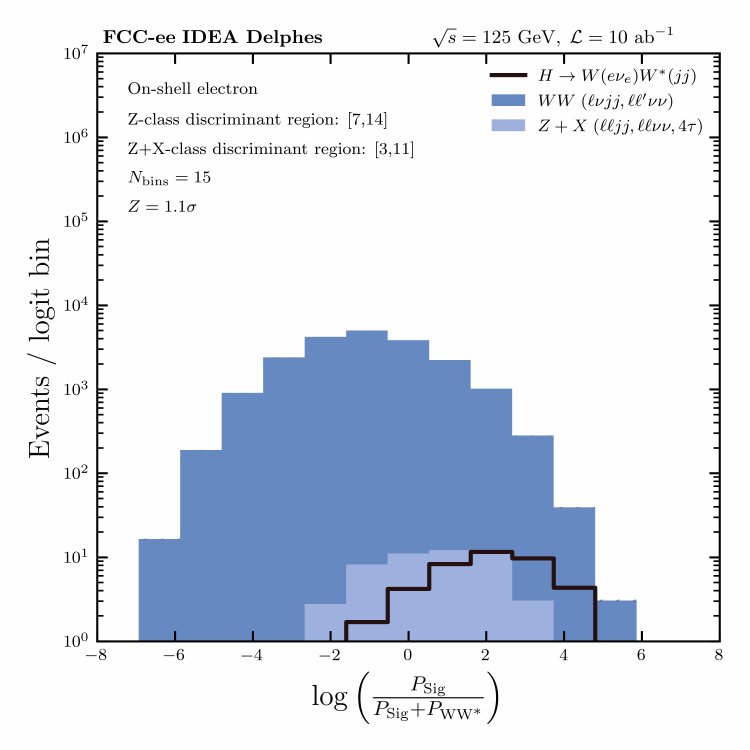}
    \includegraphics[width=0.44\textwidth]{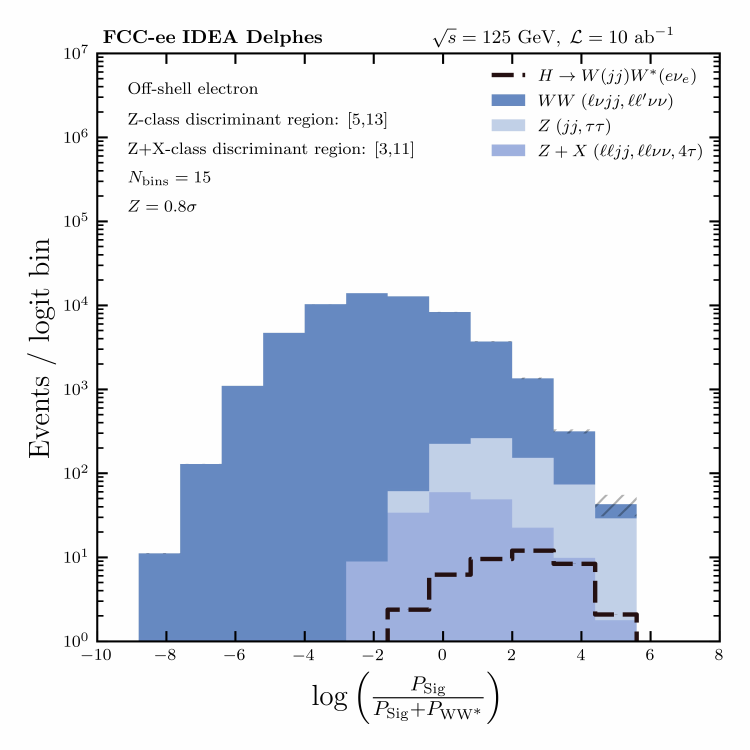}
    \includegraphics[width=0.44\textwidth]{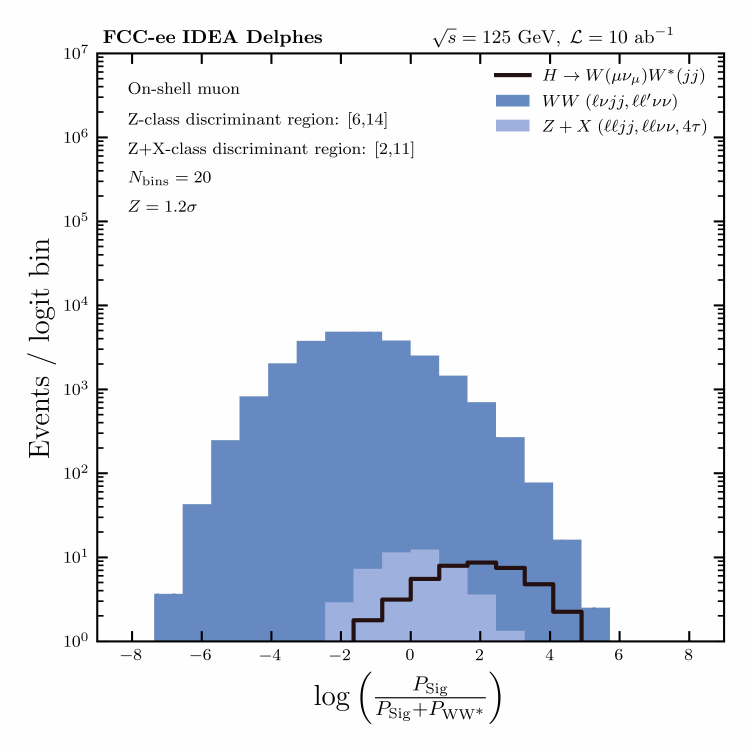}
    \includegraphics[width=0.44\textwidth]{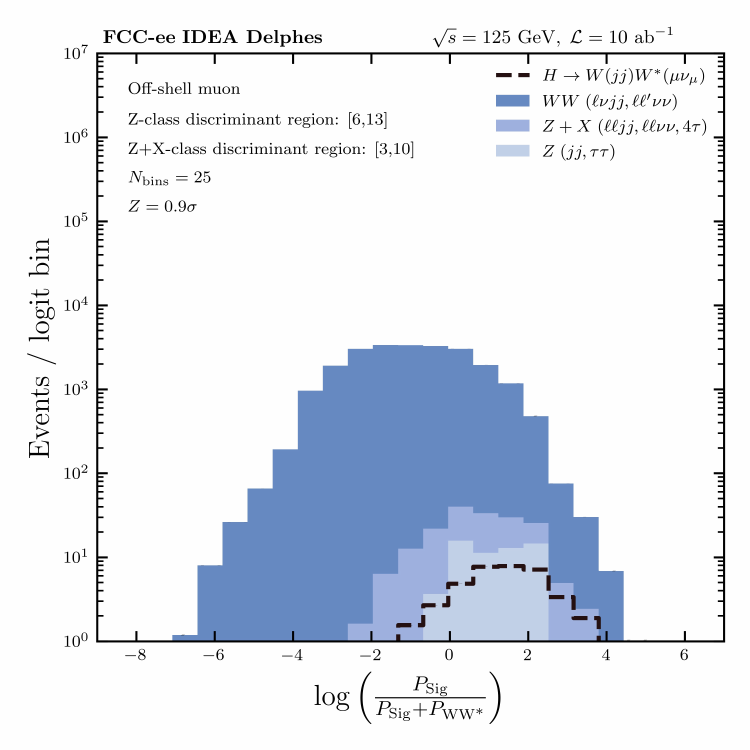}
\caption{\small Distributions of the GBDT binary discriminants (with bin counts indicated) for the signal (black lines) and the backgrounds (colored stacked histograms), after the application of the $Z^*$ and $Z+X$ binary discriminants, for the on-shell electron (upper left), off-shell electron (upper right),  on-shell muon (lower left), and off-shell muon (lower right) categories. The hatched bands (barely visible in the rightmost bins) represent the statistical uncertainty of the summed background prediction in each bin.
\label{fig:binary_W}}
\end{figure}

The binary discriminant distributions for signal and background processes, weighted by their corresponding cross sections and integrated luminosity of 10\,ab$^{-1}$, are binned into histograms spanning the full range of observed discriminant values across all event categories passing the $Z^*$ and $Z+X$ background cuts.
The selected events after the GBDT application over these reducible binary discriminants consist of signal and a manageable number of remaining $WW^*$ continuum events, which represent the most challenging background. The distributions of the GBDT binary discriminants are shown in Fig.~\ref{fig:binary_W} for the signal (solid lines) and the remaining backgrounds (colored histograms) for the four categories. 
The hatched band over the stacked background distributions represents the statistical uncertainty of the total background prediction in each bin, obtained by combining the statistical uncertainties of the individual MC background samples.
These same events are then used to construct the $WW^*$ binary discriminant, which, together with the signal classifier output, serves as input for a one-dimensional binned likelihood fit, enabling an optimal final extraction of the signal over the background.

Pseudo-data samples are constructed by summing the binned distributions of the discriminant variables for all signal and background processes, each scaled to their SM cross sections times the assumed integrated luminosity 10\,ab$^{-1}$. The statistical interpretation is based on a binned likelihood function built as the product of Poisson probabilities over all bins of the $WW^*$ binary discriminant. In each bin, the observed pseudo-data count is compared to the expected yield from the sum of signal and background contributions. The signal yield in each bin is parameterized as the product of the SM expectation and a signal strength modifier. 
The sideband regions of the discriminant distributions provide sufficient constraints on the normalizations of the main background processes. These normalizations are left floating freely in the fit, as tests showed that constraining them with a prior at the few percent level yields consistent results.

Uncertainties from the finite size of simulated samples are incorporated using the Beeston--Barlow method~\cite{Barlow:1993dm}. Shape systematic uncertainties from background modeling and experimental uncertainties related to event selection and flavor tagging are not included at this stage. The large control samples that will be available at the FCC-ee, together with expected improvements in theoretical calculations~\cite{Blondel:2019vdq} over the next two decades, are anticipated to constrain the shapes and normalizations of these processes at the permil level or better~\cite{FCC:2025lpp}.

The individual channel significances obtained from the binned likelihood fit are
$1.1$\,s.d.\ and $0.8$\,s.d.\ for the on- and off-shell electron channels,
and $1.2$\,s.d.\ and $0.9$\,s.d.\ for the on- and off-shell muon channels,
respectively (Table~\ref{tab:significance_fit}). The on-shell channels exhibit
slightly better sensitivity, as expected from the more favorable signal to
background ratio. All four channels contribute
comparably, and yield to a combined significance of $2.0\,$s.d.\ for the
$e^+e^- \to H \to WW^* \to \ell\nu + jj$ process within the
benchmark scenario of $\sigma_{e^+e^- \to H} = 280\,\mathrm{ab}$ and
$\mathcal{L}_{\mathrm{int}} = 10\,\mathrm{ab}^{-1}$. 
The statistical significance corresponding to any other alternative running scenario in the $(\delta_{\sqrts},\LumiInt)$ plane
can be derived by exploiting the bidimensional map of Ref.~\cite{dEnterria:2021xij}. 

\begin{table}[htbp!]
\centering
\caption{\small Statistical significance (in standard-deviation units) per channel after the binned likelihood fit performed over the $WW^*$ binary discriminant. The last row indicates the combined significance of all channels.}
\label{tab:significance_fit}
\renewcommand{\arraystretch}{1.2} 
\setlength{\tabcolsep}{18pt}
\resizebox{\textwidth}{!}{%

\begin{tabular}{lcccc}
\hline\hline
 & on-shell e & off-shell e & on-shell $\mu$ & off-shell $\mu$ \\
\hline
Significance [s.d.] & 1.1 & 0.8 & 1.2 & 0.9 
\\
Total Significance [s.d.] & \multicolumn{4}{c}{2.0} 
\\
\hline\hline
\end{tabular}
}
\end{table}

The factor of four gain in sensitivity with respect to the previous generator-level study of
Ref.~\cite{dEnterria:2021xij}, which reported $0.5\,$s.d.\ for the $H \to WW^*$ semileptonic channel, and a combined final statistical significance of 1.3\,s.d.\ over different Higgs decay final states, 
is driven by several factors. First, the separate treatment of the on- and off-shell $W$ boson decay
channels, combined with the independent categorization of electron and muon final
states, allows the analysis to exploit kinematic features that are diluted
in an inclusive approach. Second, our present classifier uses a better 
optimized model setup for the four separate categories, and contains more input BDT variables than the inclusive TMVA study of Ref.~\cite{dEnterria:2021xij}. The inclusion of, \eg\ jet flavor-tagging observables among the discriminating variables also plays a particularly important role in this improved separation. Third, the multiclass GBDT classifier, trained simultaneously against the $WW^*$ continuum, $Z+X$, and $Z^*$ backgrounds, effectively eliminates the reducible contributions and leaves a manageable level of irreducible $WW^*$ continuum background for the final likelihood fit. If we switch back our GBDT model to a single-class (with all backgrounds combined, as in the study of Ref.~\cite{dEnterria:2021xij}, we obtain about twice smaller significances than listed in Table~\ref{tab:significance_fit} because the classifier exploits less well the properties of the various backgrounds with different kinematics. These results also suggest that additional signal $WW^*$ decay modes not considered here, such as channels involving hadronic $\tau$ decays, the fully leptonic $2\ell 2\nu$, and the fully hadronic $4j$ final states, may further improve the overall sensitivity~\cite{InPrep}.

A cross section excess with 2\,s.d.\ significance is equivalent to a (Gaussian) signal strength of $\mu = 1\pm 0.5$ (1\,s.d.).
For a one-parameter Gaussian likelihood, 95\% CL (one-sided) corresponds approximately to $+1.64\,$s.d., and then $\mu_{95}\approx 1+1.64 \times 0.5=1.82$.
Since the Higgs dielectron decay width depends on the square of the Yukawa coupling, $\Gamma(H\to\epem)\propto y_e^2$, our expected 2\,s.d.\ excess
translates into an upper bound on the electron Yukawa coupling modifier of $\kappa_e \lesssim \sqrt{1.82} = 1.35$ at 95\% CL. 
If, instead, no excess would be observed in the data after performing the measurement outlined here, \ie\ if we found
a result consistent with the background-only assumption, the limit would be $\mu_{95}\approx 1.64 \times 0.5=0.82$, which would therefore 
translate into $\kappa_e \lesssim \sqrt{0.82} = 0.9$ at 95\% CL.


\section{Summary} \label{sec:Conclusion}

This work has presented a feasibility study for the observation of resonant $s$-channel Higgs boson production in $\epem$ collisions at $\sqrts = 125$\,GeV at the FCC-ee, focusing on the $WW^*$ decay mode in the lepton+jets final states, namely $e^+e^- \to H \to WW^* \to \ell\nu + jj$. The analysis is performed considering a signal production cross section at the Higgs pole of $\sigma_{e^+e^- \to H} = 280\,\mathrm{ab}$, accounting for the effects of initial state radiation (ISR) and a beam monochromatization leading to a center-of-mass spread of $\delta_{\sqrts}\approx 4.1$\,MeV commensurate with the natural Higgs boson width, and an integrated luminosity of $10\,\mathrm{ab}^{-1}$. Signal events are categorized according to the on-shell and off-shell nature of the $W$ bosons and their corresponding leptonic and hadronic decay modes. Separate analyses are performed for electron and muon final states, with the leptonic $\tau$ decays included and combined with the corresponding $e$ and $\mu$ channels. In total, four signal categories are considered, together with their multiple background contributions arising from $Z^*$, $Z+X$, and continuum $WW^*$ processes.

To enhance the separation between signal and background, a multi-variate-analysis using a multiclass gradient boosted-decision-tree (GBDT) classification approach is employed. The GBDT classifier employs 95 kinematic and topological input features to separate signal from background processes. Four event classes are defined, corresponding to the signal and the three main background categories. The application of dedicated multi-class binary discriminants significantly suppresses the reducible background contributions, resulting in a background composition dominated by the irreducible $WW^*$ continuum. The final signal sensitivity is extracted using a one-dimensional binned likelihood fit to the $WW^*$ binary discriminant. For the $H \to WW^* \to \ell\nu + 2\,\mathrm{jets}$ final state only, the combined expected statistical significance (excluding hadronic $\tau$ decays) reaches approximately $2.0$\,standard-deviations, representing an improvement of a factor of four over the previous generator-level study for this decay channel~\cite{dEnterria:2021xij}. This translates into an upper bound on the electron Yukawa coupling modifier of $\kappa_e = y_e/y_e^{\rm \textsc{sm}} \lesssim 1.35$ at 95\% confidence level for this Higgs decay mode alone, and represents the most stringent experimental constraint on the electron Yukawa coupling obtained from simulation studies to date.



\section*{Acknowledgments}

R.\ Jafari and K.\ Azizi acknowledge the financial support provided by the Iran National Science Foundation (INSF) under grant number 4040738. R.\ Jafari and A.\ Fatehi thank CERN for the kind hospitality and financial assistance.


\bibliographystyle{JHEP}
\bibliography{main}


\appendix

\section{Cone isolation optimization}\label{App:A}

\begin{figure}[htpb!]
    \centering
    \includegraphics[width=0.99\textwidth]{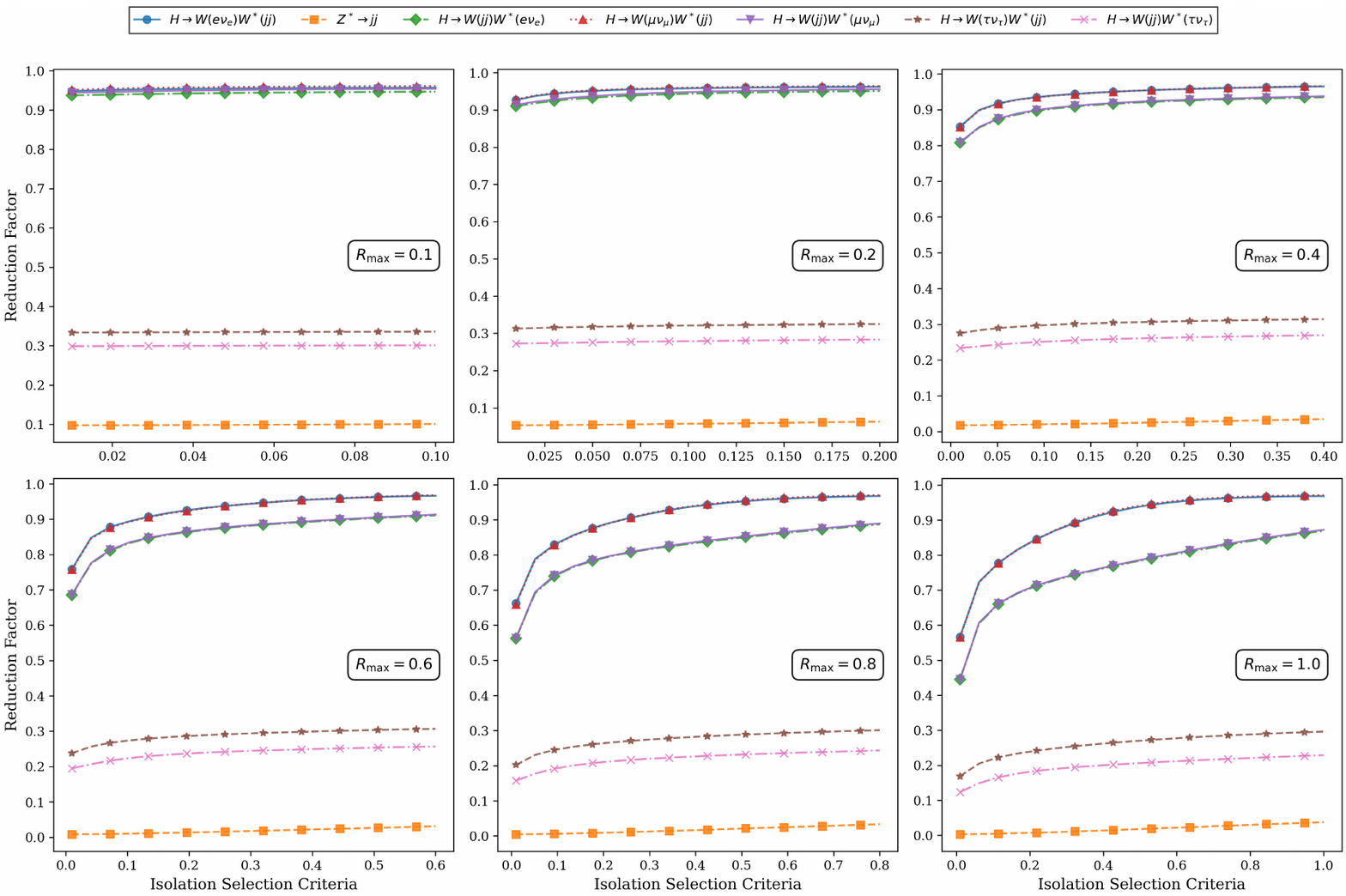}
    \caption{Optimization of the lepton isolation cone radius, $R$, and relative hadronic momentum fraction inside the cone, $I_\mathrm{rel}$, for various signal and background final states (symbols with curves). The $y$ axes show the reduction factor in the signal and background yields as a function of the $I_\mathrm{rel}$ parameter ($x$ axis), for varying values of the cone radius $R$.}
    \label{plot:separateDrMax}
\end{figure}

The charged-lepton isolation parameters, including the cone radius ($R$) and the relative charged-hadron momentum fraction inside the cone ($I_\mathrm{rel}$) variables, have been optimized in this analysis following the results shown in Fig.~\ref{plot:separateDrMax}.
The selection of a cone of radius $R = 0.1, 0.2$ retains the same and maximum possible signal events with the corresponding isolation cut of $I_\mathrm{rel} = 0.1 , 0.2$. 
However, for the a cone of radius $R=0.2$, more ISR events are removed. Therefore, the cone radius of $R=0.2$ and isolation fraction of $I_\mathrm{rel}=0.2$ are finally selected. 

\section{MVA input variables, binary discriminants, and GBDT network configuration}\label{App:B}

\begin{figure}[htpb!] 
\centering
\includegraphics[width=0.325\textwidth]{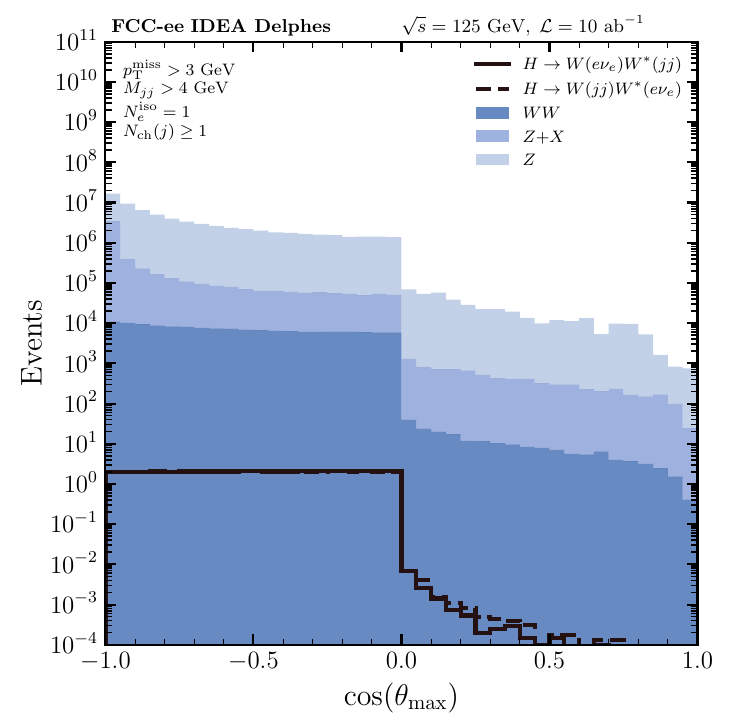}
\includegraphics[width=0.325\textwidth]{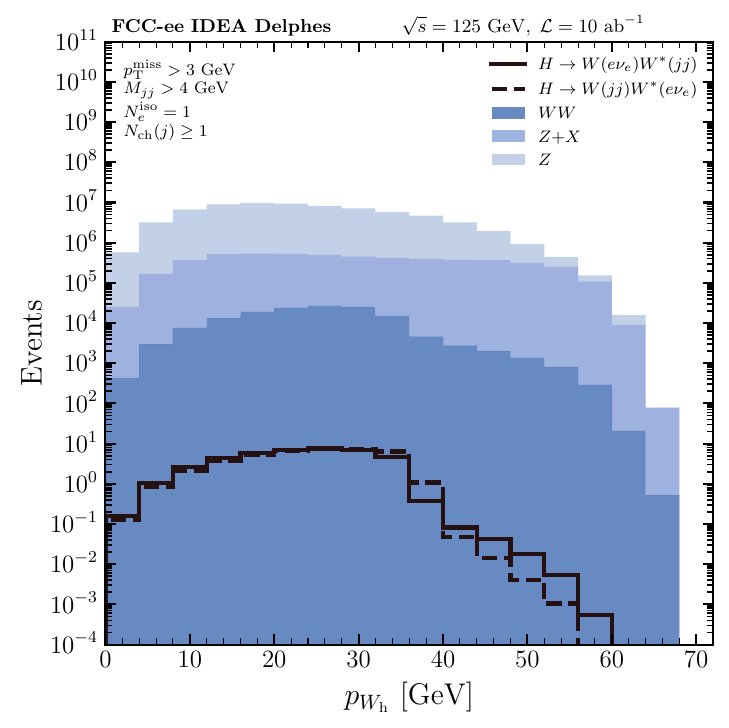}
\includegraphics[width=0.325\textwidth]{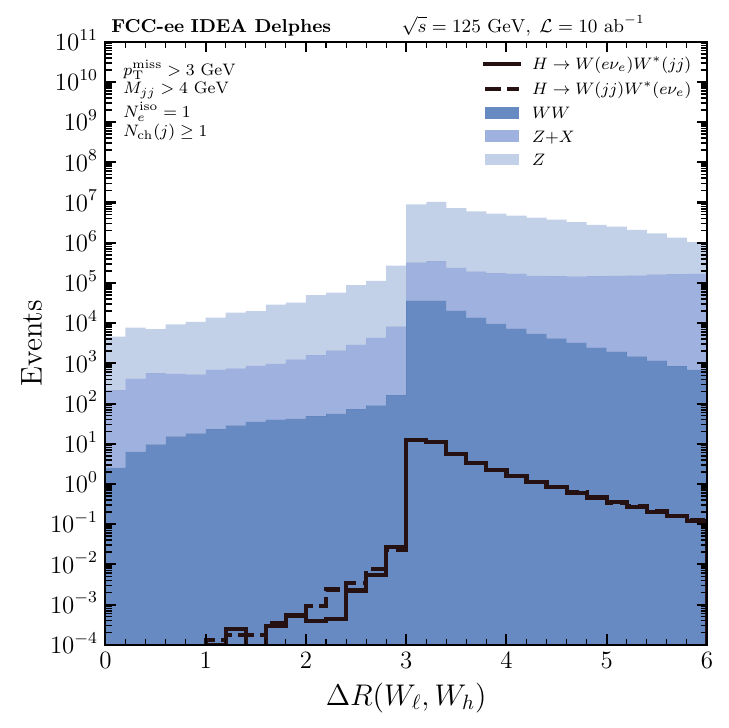}
\includegraphics[width=0.325\textwidth]{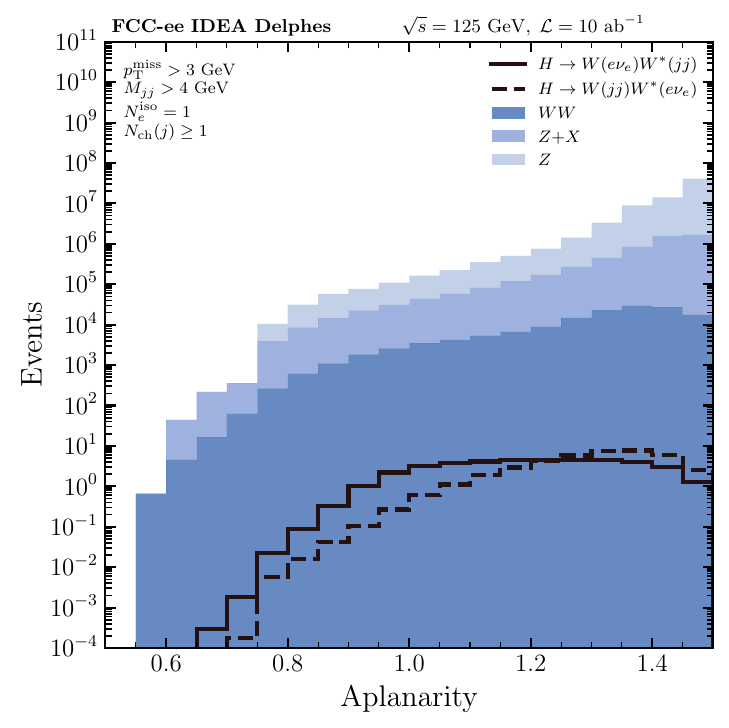}
\includegraphics[width=0.325\textwidth]{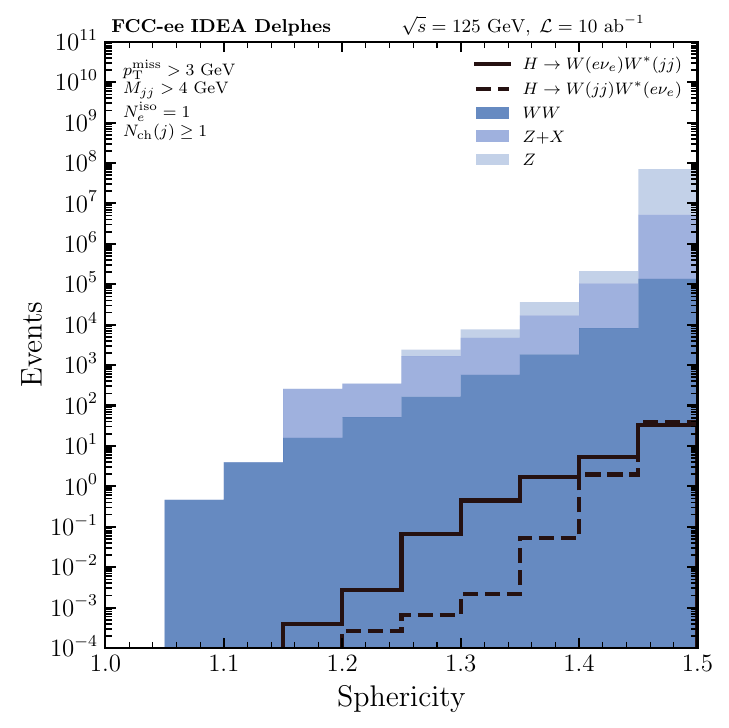}
\includegraphics[width=0.325\textwidth]{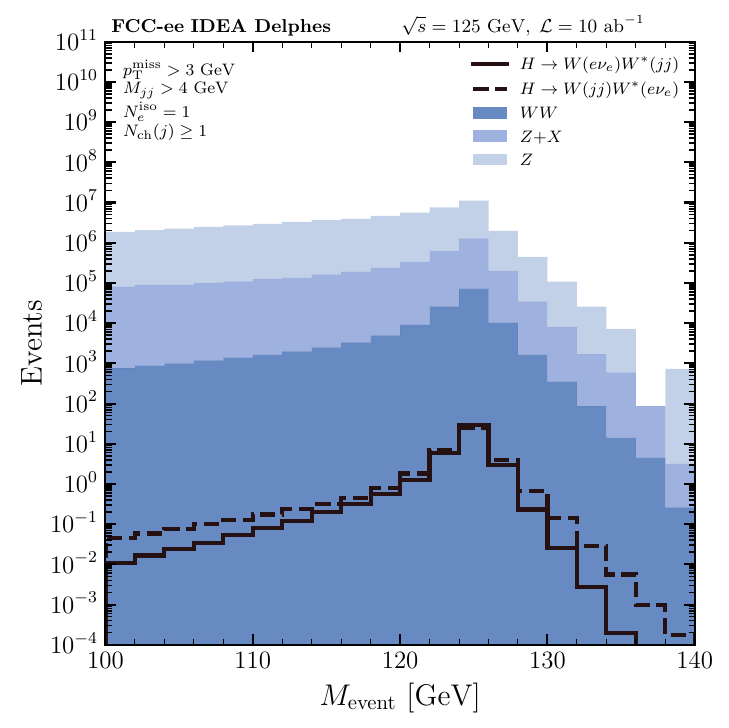}
\includegraphics[width=0.325\textwidth]{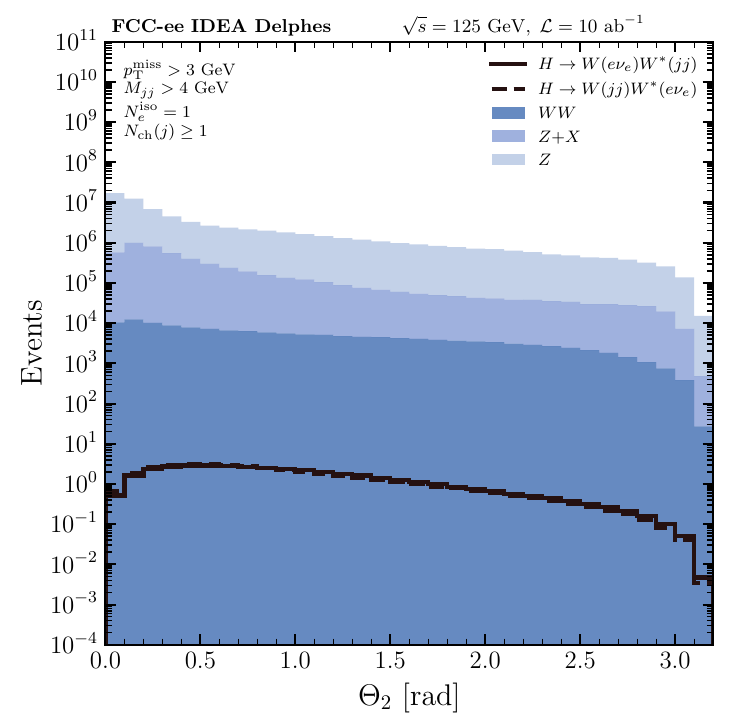}
\includegraphics[width=0.325\textwidth]{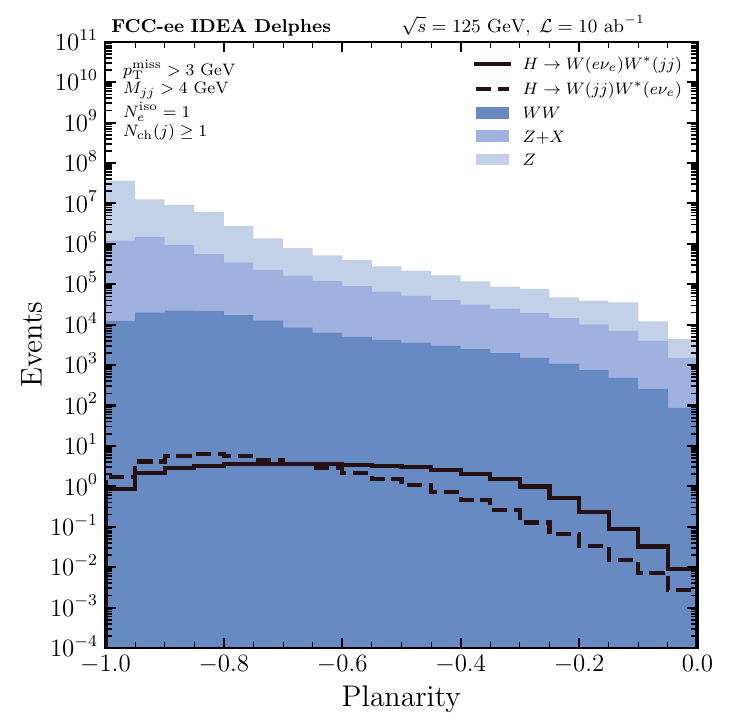}
\includegraphics[width=0.325\textwidth]{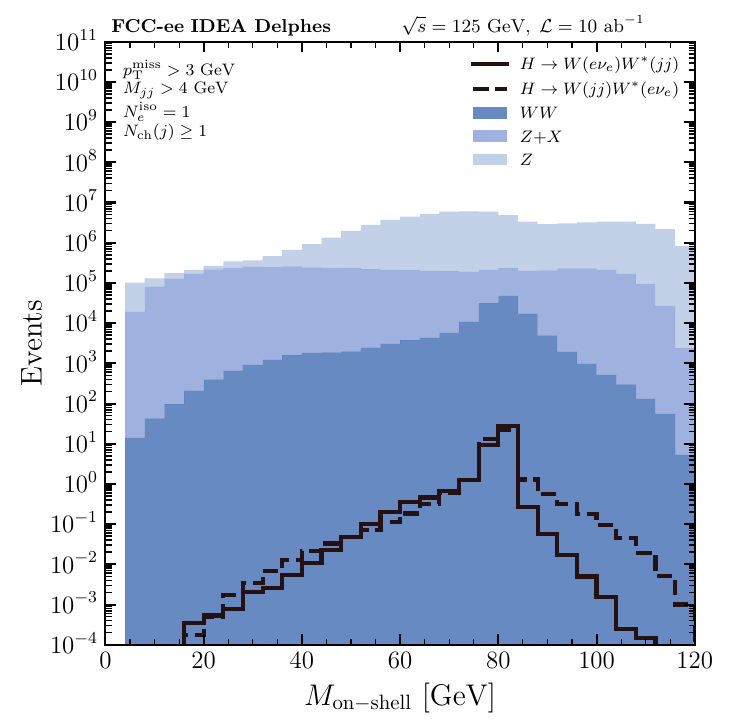}
\includegraphics[width=0.325\textwidth]{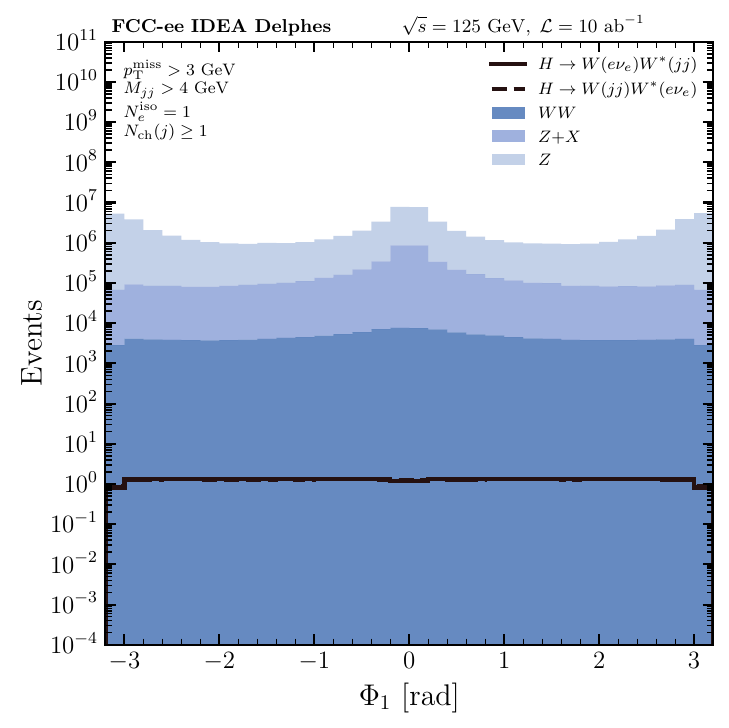} 
\includegraphics[width=0.325\textwidth]{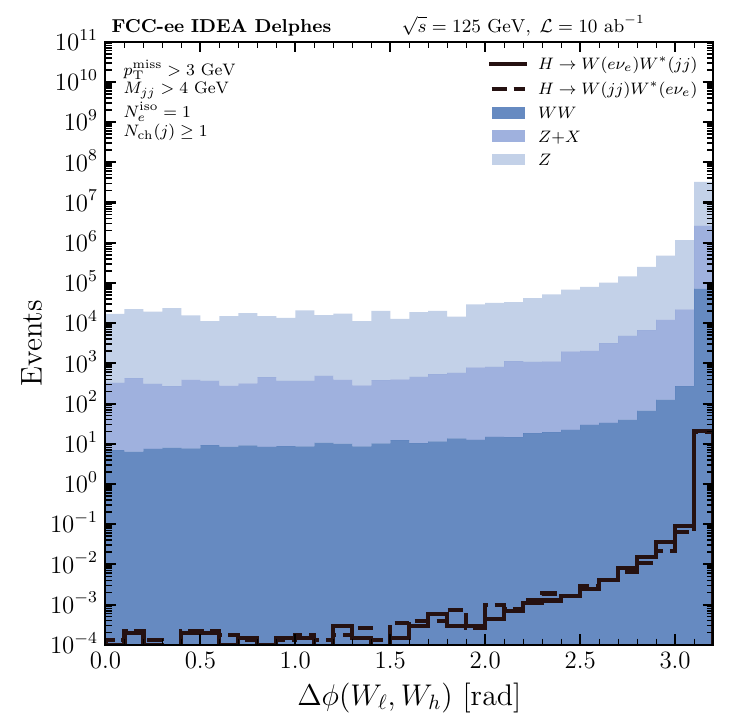}
\includegraphics[width=0.325\textwidth]{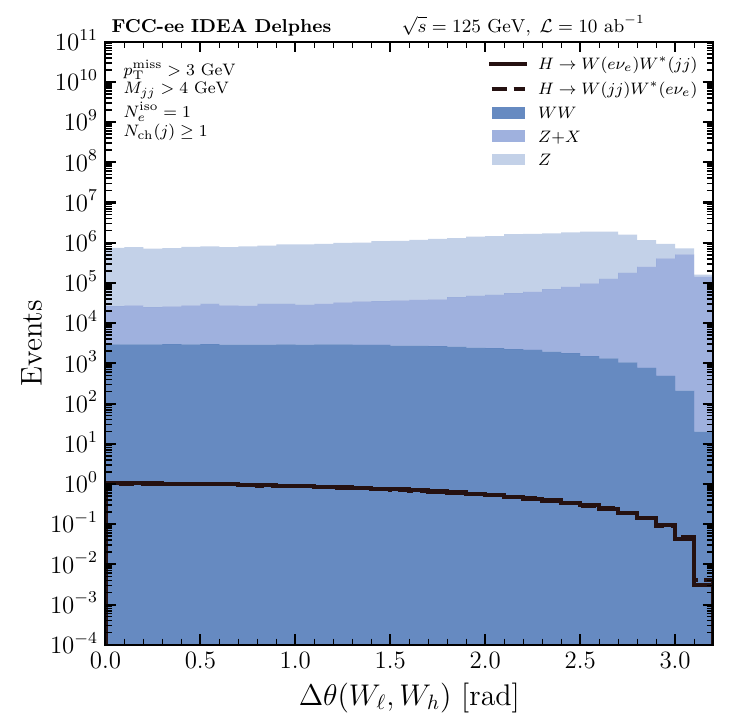}
\caption{\small {Examples of GBDT input variable distributions used in this analysis in the electron final states for the Higgs signal (solid and dashed black lines, for the on- and off-shell channels, respectively) 
and backgrounds (colored stacked histograms) in $\epem$ collisions at $\sqrt{s}=125$\,GeV. The number of events corresponds to $\LumiInt = 10$\,ab$^{-1}$.}
\label{fig:MVA_inputs}}
\end{figure}

\begin{figure}[htpb!]
\centering
    \includegraphics[width=0.32\textwidth]{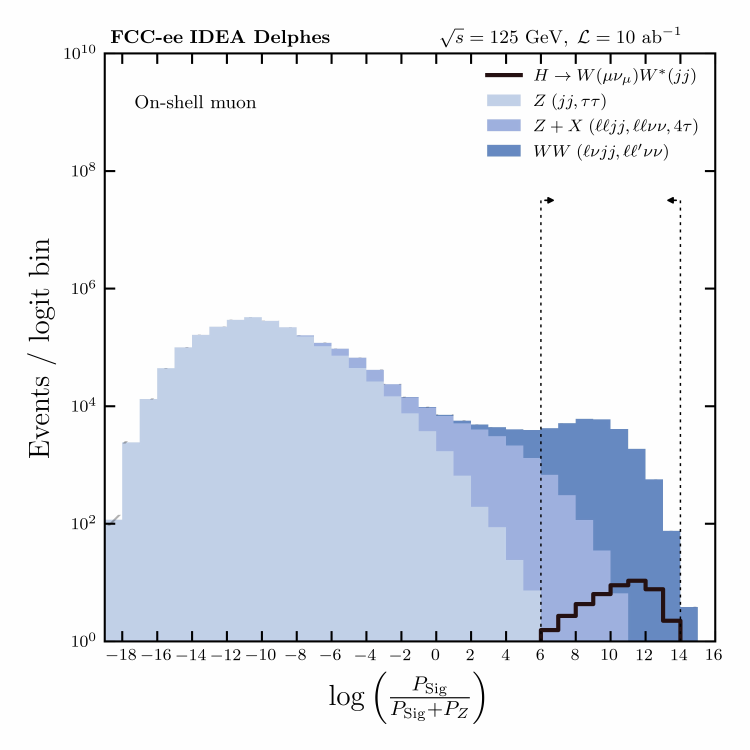}
    \includegraphics[width=0.32\textwidth]{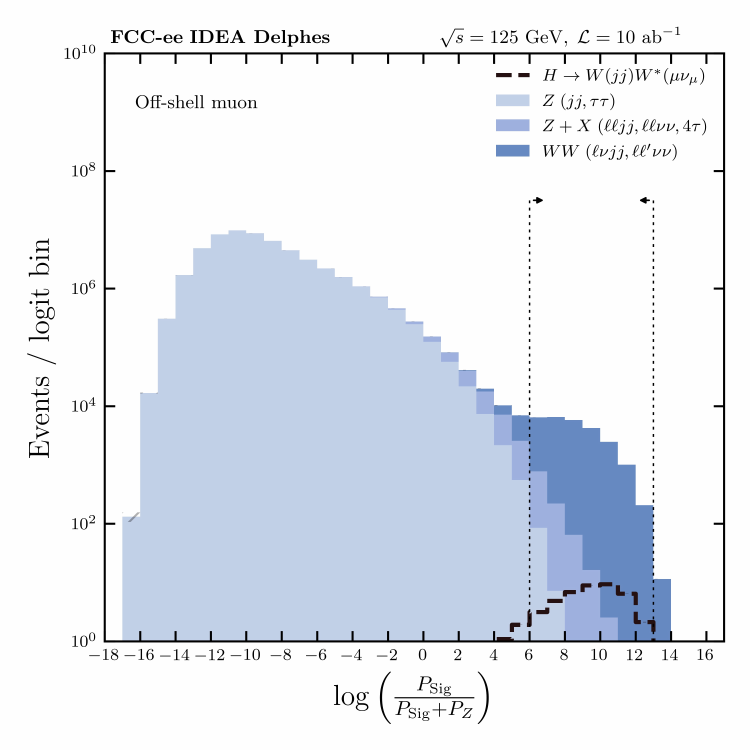}
    \includegraphics[width=0.32\textwidth]{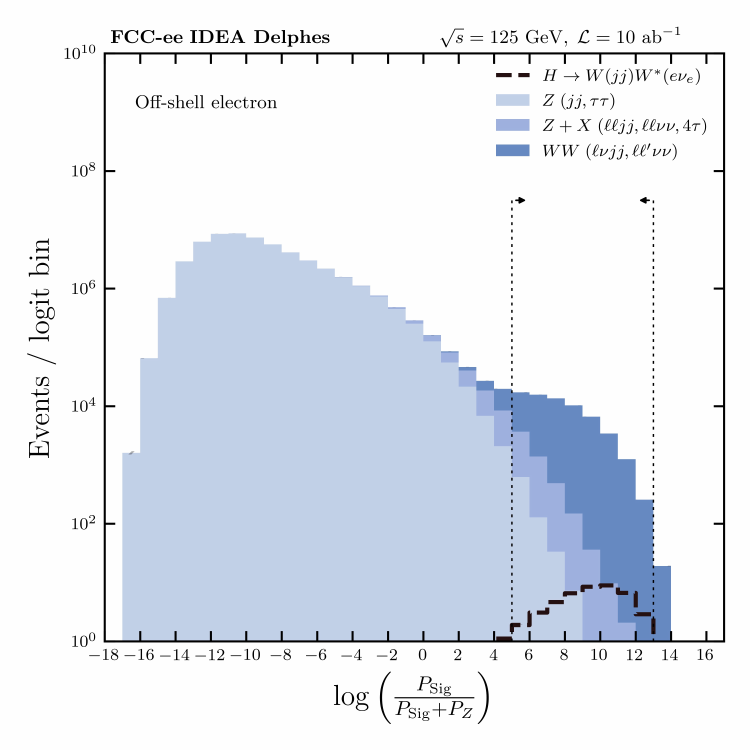}
    \includegraphics[width=0.32\textwidth]{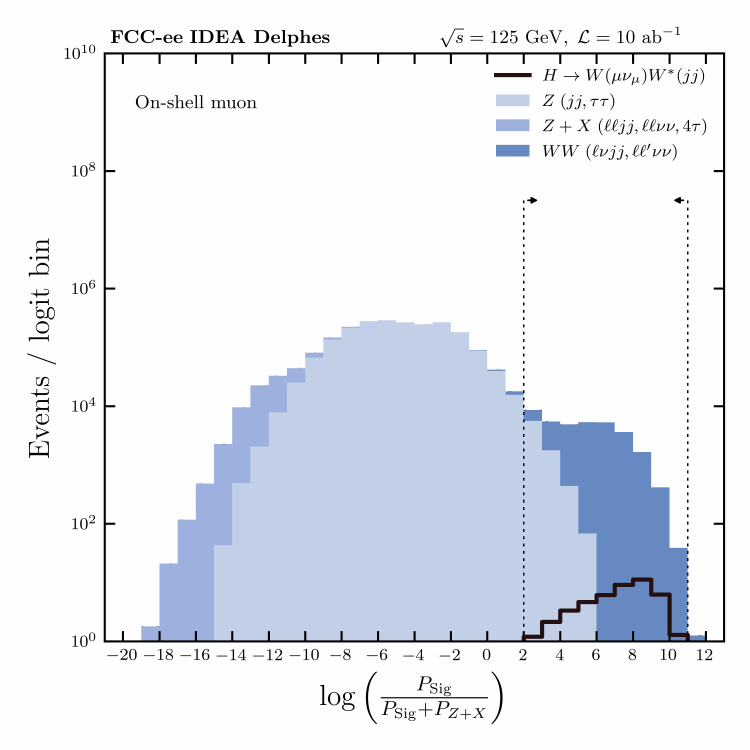}
    \includegraphics[width=0.32\textwidth]{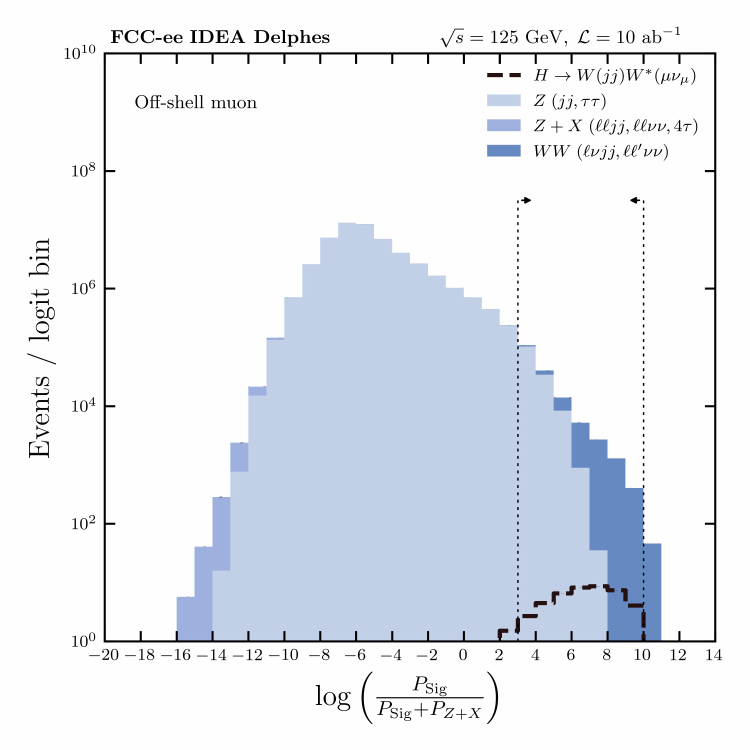}
    \includegraphics[width=0.32\textwidth]{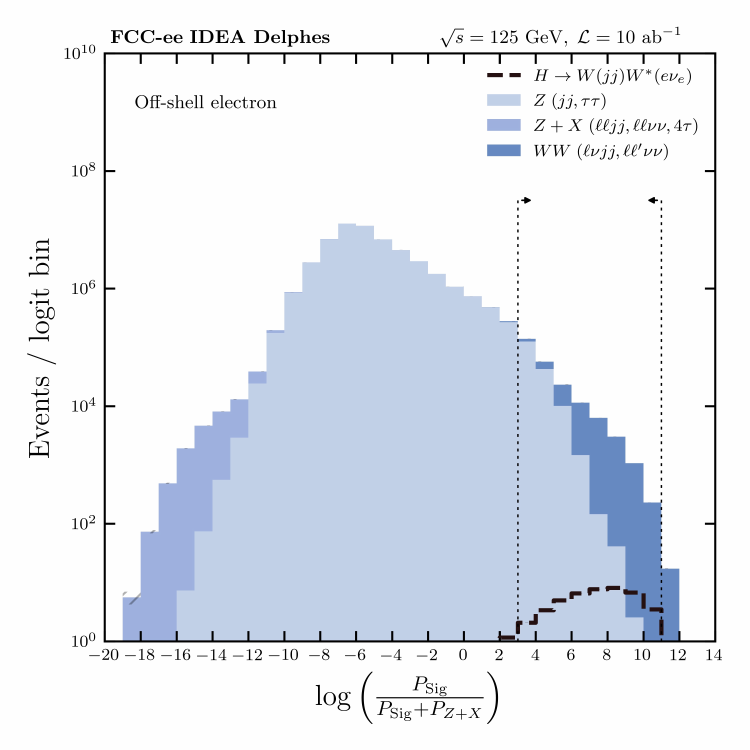}
\caption{\small Distributions of the GBDT binary discriminants (with the bin counts shown) for signal (solid or dashed black lines) and $Z^*$, $Z+X$, and $WW^*$ backgrounds (colored histograms), before applying cuts on the $Z^*$ (upper) and $Z+X$ (lower) binary discriminants for the  on-shell muon (left), off-shell muon (center), and off-shell electron (right) categories (the corresponding on-shell electron distributions can be found in Fig.~\ref{fig:binary}). The vertical dashed lines indicate the signal-dominated ranges where the cuts are applied for the removal of $Z^*$ and $Z+X$ backgrounds. \label{fig:other_binary}}
\end{figure}

Table~\ref{tab:BDT_full_vars} presents a fraction of the MVA input variables used in our analysis, ranked in order of approximate discriminating power, for the on-shell electron channel.
The distributions of a few representative variables are shown in Fig.~\ref{fig:MVA_inputs}, for the on-shell electron channel.
%
%
Additionally, Fig.~\ref{fig:other_binary} displays the signal--background binary discriminants for the $Z+X$ and $Z^*$ reducible backgrounds before the final selection cuts. The optimized intervals on these GBDT discriminants to maximize the signal separation from backgrounds are: 
\begin{itemize}
\item Off-shell electron: $[5,13]$ (for $Z^*$), and $[3,11]$ (for $Z+X$)
\item On-shell muon: $[6,14]$ (for $Z^*$), and $[2,11]$ (for $Z+X$)
\item Off-shell muon: $[6,13]$ (for $Z^*$), and $[3,10]$ (for $Z+X$)
\end{itemize}
Furthermore, Table~\ref{tab:network_config} summarizes the complete configuration of the \textsc{XGBoost} multiclass classifier, including network architecture, class definitions, tree structure parameters, learning configuration, and data preprocessing settings employed in this analysis.

\begin{table}[htbp!]
\caption{\small Representative set of 64 GBDT input variables employed in the on-shell electron channel, with a ranking indicative of their approximate discrimination importance.
\label{tab:BDT_full_vars}}
\centering
\setlength{\tabcolsep}{12pt}
\small
\resizebox{\textwidth}{!}{%

\begin{tabular}{@{}ll|ll|ll|ll@{}}
\hline\hline
Rank & Variable & Rank & Variable & Rank & Variable & Rank & Variable \\
\hline
1  & $\mathcal{P}_{b\bar{b}}$                  & 17 & $\Delta M(W,W^*)$                & 33 & $\phi_{W,W^*}^\mathrm{max}$      & 49 & $M_W^\mathrm{onshell}$ \\
2  & $\cos\theta_{W,W^*}^\mathrm{max}$         & 18 & $\mathcal{P}_{c\bar{c}}$         & 34 & $\cos{\theta_{j_1}}$             & 50 & $M_{j_2}$ \\
3  & $\mathcal{P}_{s\bar{s}}$                  & 19 & $p_{\gamma^{\rm iso}}$           & 35 & $p_{j_2}$                        & 51 & Asphericity \\
4  & $p_{W_{had}}$                             & 20 & $\cos\theta_{j_2}$               & 36 & $\Phi_\mathrm{1,\,MELA}$             & 52 & $d_{34}$ \\
5  & $\tau$-jet flavor score                   & 21 & $p_{j_1}$                        & 37 & Planarity                        & 53 & $\mathcal{P}_{d\bar{d}}$ \\
6  & $\cos\theta_\mathrm{miss}$                & 22 & Sphericity                       & 38 & $d_{23}$                         & 54 &  $M_\ell^\mathrm{iso}$\\
7  & $M_W^\mathrm{offshell}$                   & 23 & $M_\mathrm{event}$               & 39 & $\cos\Theta^*_{\mathrm{MELA}}$   & 55 &  $\Delta \theta(W,W^*)$\\
8  & $M_{W_\ell}$                              & 24 & $p_\mathrm{miss}$                & 40 & $\mathcal{P}_{u\bar{d}}$         & 56 & $\cos\theta_{W_\ell}$ \\
9  & $\Delta R(W_\ell,W_h)$                    & 25 & $\cos\theta_{\gamma^{\rm iso}}$  & 41 & $E_{j_2}$                        & 57 & $\phi_\gamma^\mathrm{iso}$ \\
10 & $M_{W_h}$                                 & 26 & $p_{\ell^{\rm iso}}$             & 42 & $\cos\Theta_\mathrm{2,\,MELA}$        & 58 &  $E_{j_1,j_2}^\mathrm{max}$\\
11 & $\mathcal{P}_{s\bar{c}}$                  & 27 & $\cos\Theta_\mathrm{2,\,MELA}$        & 43 & $d_{34}$                         & 59 &  $M_{j_1}$\\
12 & $\angle(W_\ell,W_h)$                      & 28 & $\mathcal{P}_{u\bar{u}}$         & 44 & $\cos\theta_{W_h}$               & 60 & $N^{\rm const.}_{j_1}$ \\
13 & $\cos\theta_{W,W*}^{\rm min}$             & 29 & $\mathcal{P}_{c\bar{s}}$         & 45 & $E_{j_1,j_2}^\mathrm{min}$       & 61 & $N^{\rm const.}_{j_2}$ \\
14 & $p_{W_\ell}$                              & 30 & $E_\gamma^{\rm iso}$             & 46 & $\mathcal{P}_{d\bar{u}}$         & 62 & $E_\mathrm{miss}$ \\
15 & $\Phi_\mathrm{MELA}$                      & 31 & $\phi_{W,W^*}^\mathrm{min}$      & 47 & $\Phi^{*}_\mathrm{MELA}$         & 63 & $N_{\ell^{\rm iso}}$ \\
16 & Aplanarity                                & 32 & $N_{\gamma^{\rm iso}}$           & 48 & $\theta_{W_\ell}$                & 64 & $E_{\ell^{\rm iso}}$ \\
\hline\hline
\end{tabular}
}
\end{table}

\clearpage

\begin{figure}[htpb!] 
\centering
\includegraphics[width=0.325\textwidth]{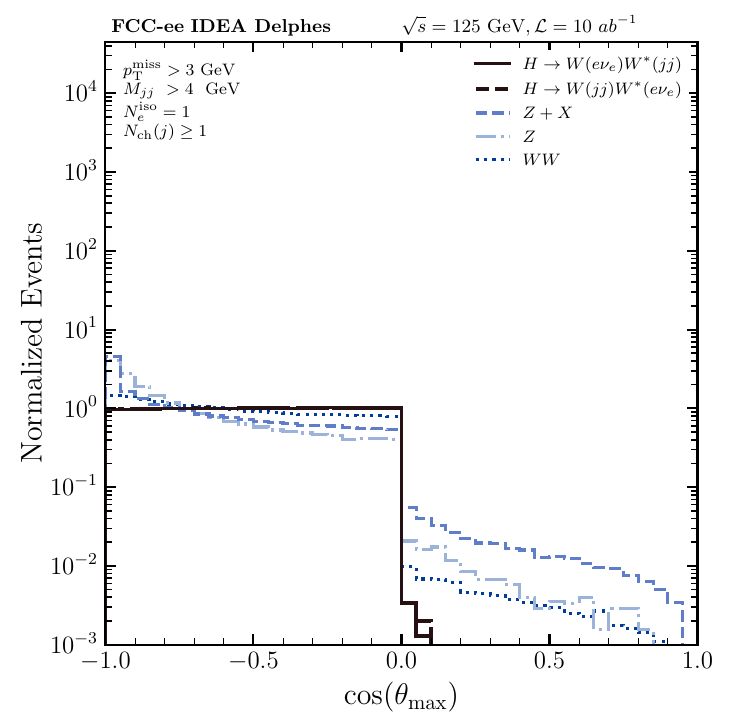}
\includegraphics[width=0.325\textwidth]{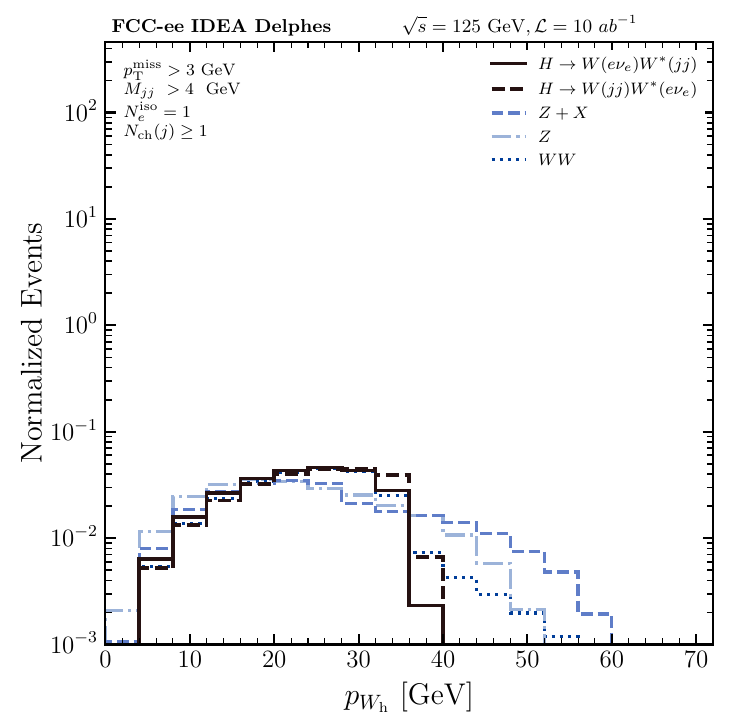}
\includegraphics[width=0.325\textwidth]{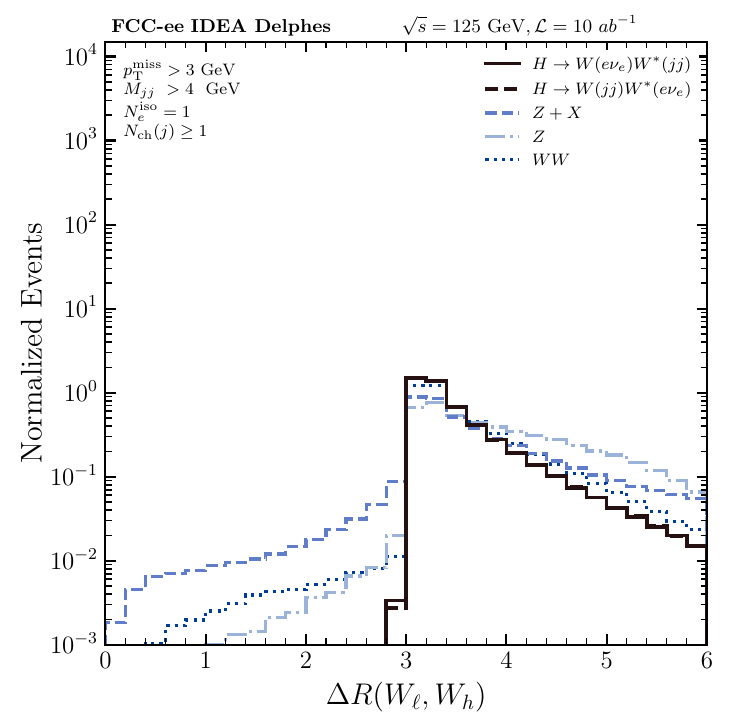}
\includegraphics[width=0.325\textwidth]{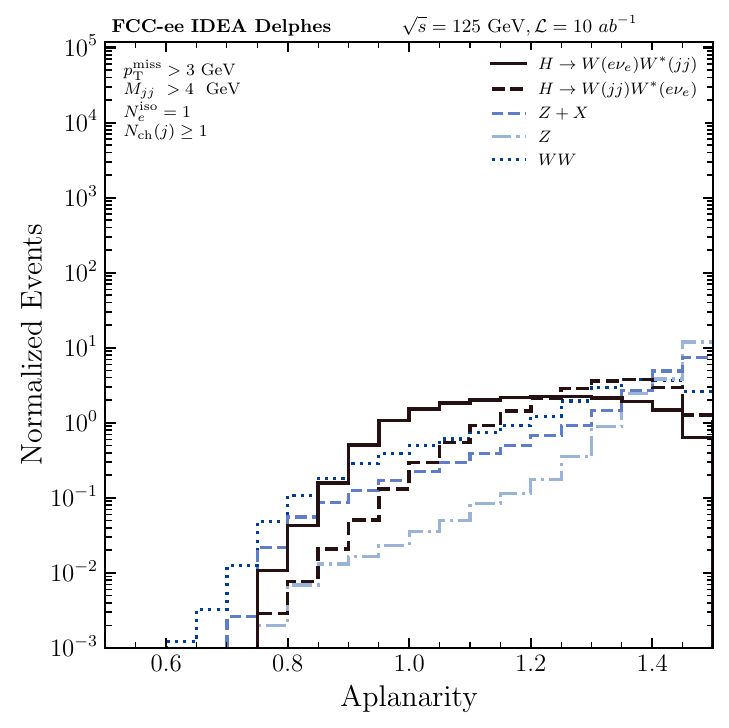}
\includegraphics[width=0.325\textwidth]{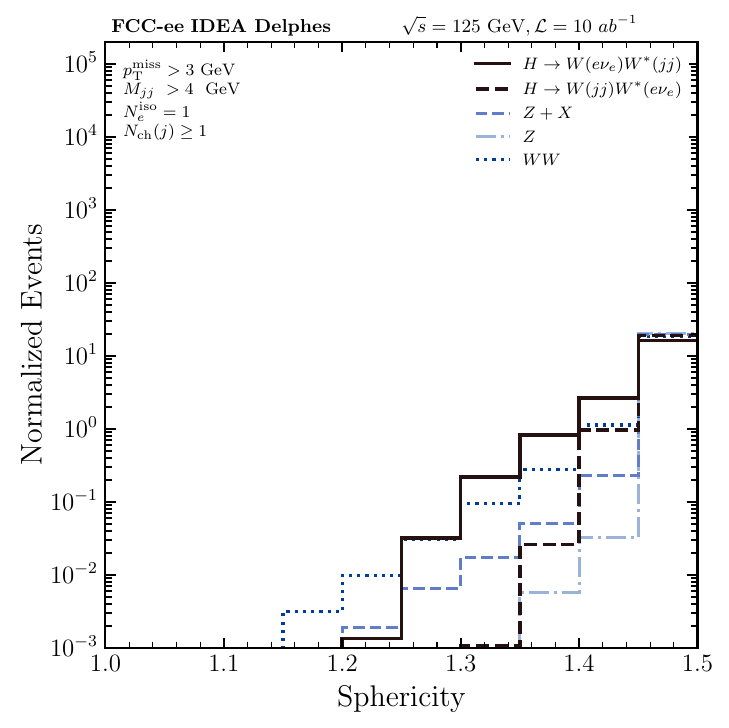}
\includegraphics[width=0.325\textwidth]{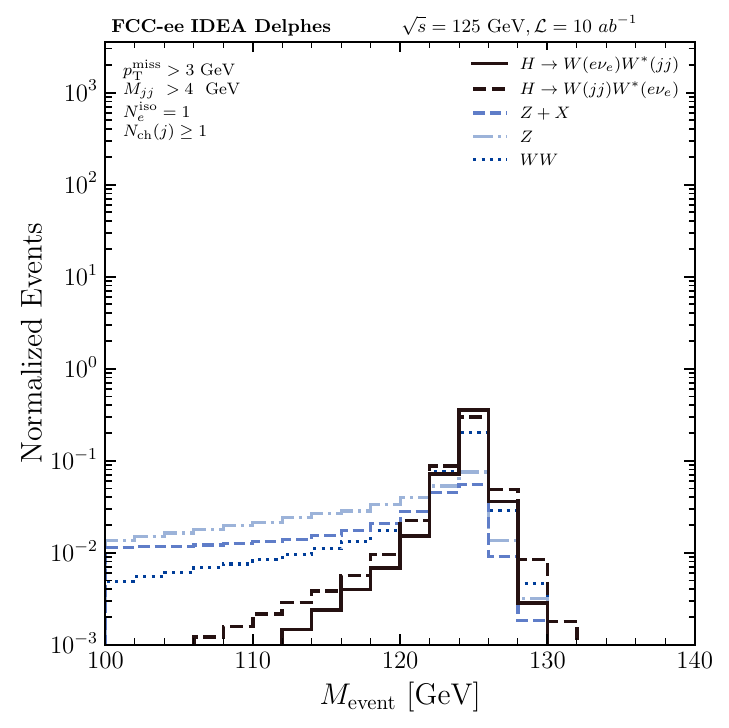}
\includegraphics[width=0.325\textwidth]{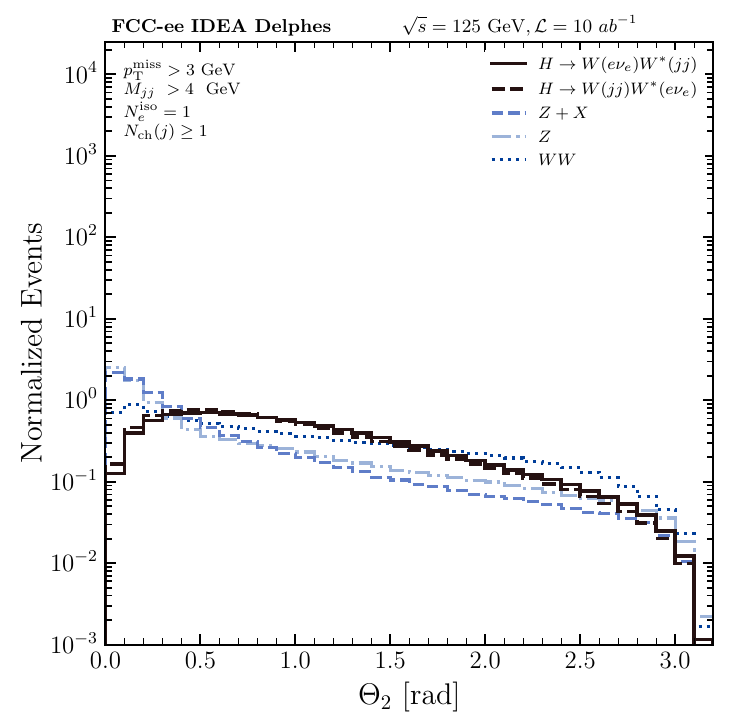}
\includegraphics[width=0.325\textwidth]{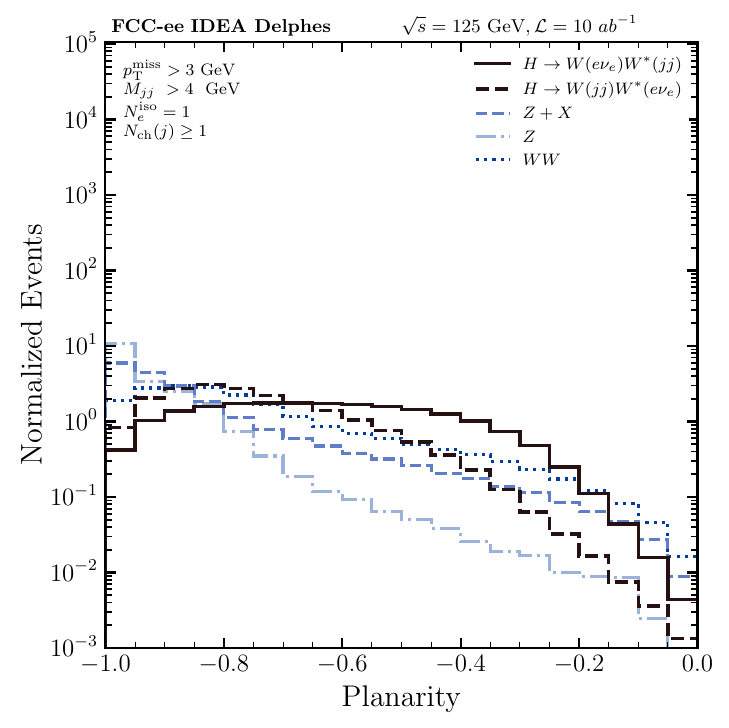}
\includegraphics[width=0.325\textwidth]{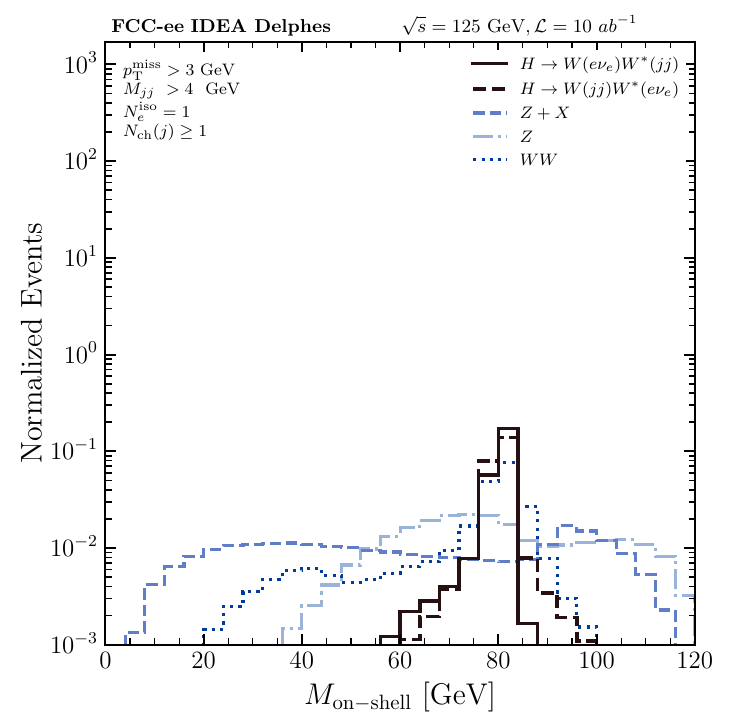}
\includegraphics[width=0.325\textwidth]{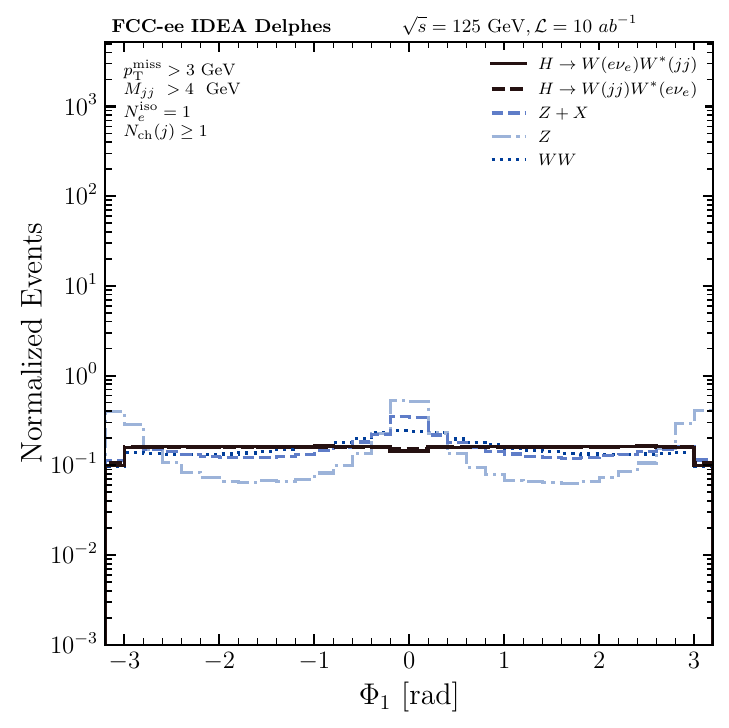}
\includegraphics[width=0.325\textwidth]{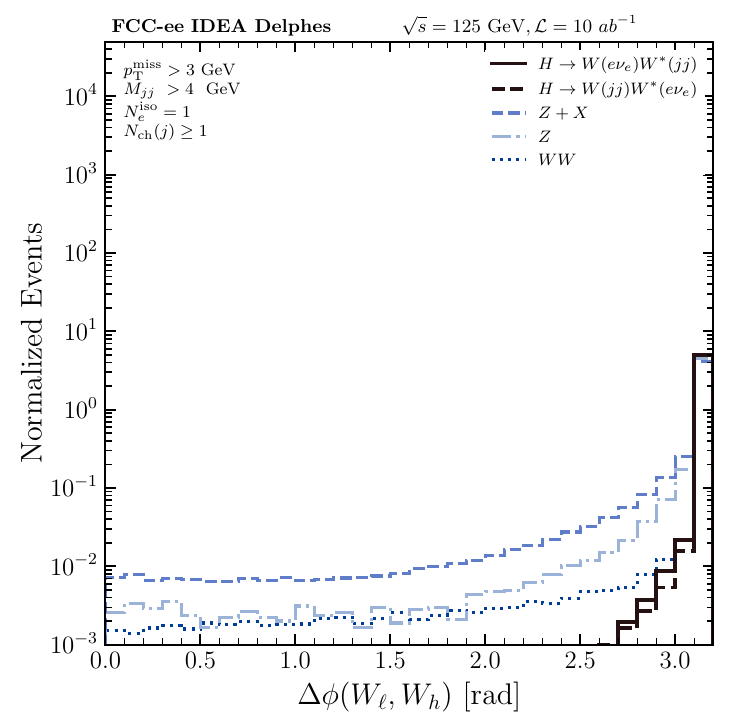}
\includegraphics[width=0.325\textwidth]{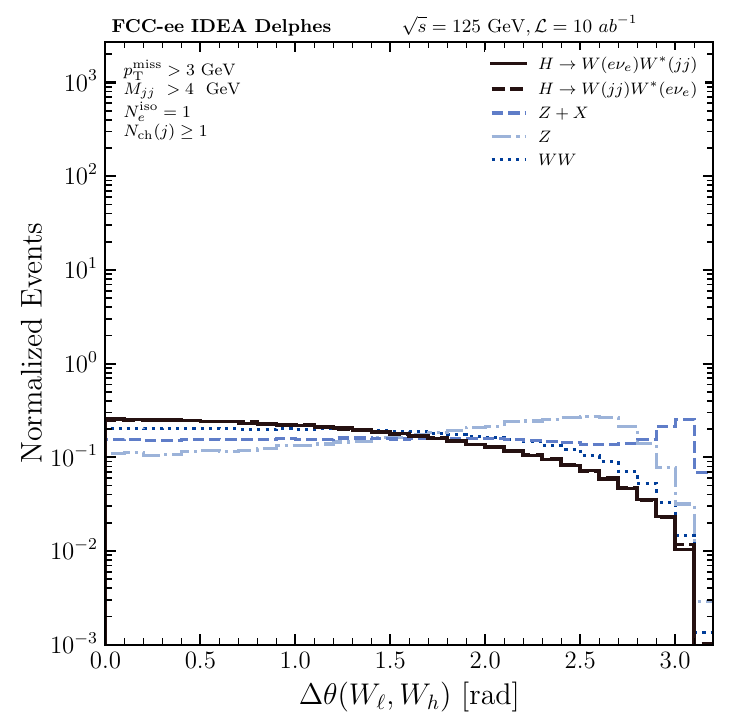}
\caption{\small Distributions of the same GBDT input variables plotted in Fig.~\ref{fig:MVA_inputs}, but normalized to unit-area for shape comparison, 
in the electron final states for the Higgs signal (solid and dashed black histograms, for the on- and off-shell channels, respectively) 
and backgrounds (dotted, dashed, and dashed-dotted blue histograms) in $\epem$ collisions at $\sqrt{s}=125$\,GeV. 
\label{fig:MVA_inputs_norm}}
\end{figure}

\begin{table}[p]
\centering
\caption{\small {\textsc{XGBoost} multiclass classification network configuration.}}
\label{tab:network_config}
\setlength{\tabcolsep}{65pt}
\resizebox{\textwidth}{!}{%

    \begin{tabular}{p{5cm}p{6cm}}
\hline\hline
\textbf{Parameter} & \textbf{Value/Description} \\
\hline
\multicolumn{2}{l}{\textbf{Network Architecture}} \\
\hline
Classification Type & multiclass classification \\
Objective Function & Softmax probability (multi:softprob) \\
Number of Output Classes & 4 \\
Number of Input Features & 95 \\
Trees per Boosting Round & 4 (one per class) \\
Total Trees (Maximum) & 120,000 (30,000 rounds $\times$ 4 classes) \\
\hline
\multicolumn{2}{l}{\textbf{Class Definitions}} \\
\hline
Class 0: Signal & $H \to WW^* \to \ell\nu_\ell jj$  \\
Class 1: $WW^*$ background & $e\nu_e jj$, $\mu\nu_\mu jj$, $\tau\nu_\tau jj$, $\ell_1\ell_2\nu\nu$ \\
Class 2: $Z+X$ background & $\tau\tau jj$, $\mu\mu jj$, $eejj$, $4 \tau$, $\tau\tau\nu\nu$ \\
Class 3: $Z^*$ background & $\tau\tau$, $jj$ \\
\hline
\multicolumn{2}{l}{\textbf{Tree Structure}} \\
\hline
Maximum Tree Depth & 6 levels \\
Maximum Leaf Nodes & 64 (= $2^6$) \\
Tree Construction Method & Histogram-based (exact splits) \\
Minimum Child Weight & 1 (default) \\
Feature Sampling per Tree & All 95 features available \\
Row Sampling per Tree & 100\% (no subsampling) \\
\hline
\multicolumn{2}{l}{\textbf{Learning Configuration}} \\
\hline
Learning Rate ($\eta$) & 0.001 (conservative update step) \\
Number of Boosting Rounds & Up to 30,000 (early stopping enabled) \\
Early Stopping Patience & 100 rounds without improvement \\
Regularization & L2 regularization (default $\lambda = 1$) \\
Loss Function & multiclass logarithmic loss \\
Gradient Information & First and second-order derivatives \\
\hline
\multicolumn{2}{l}{\textbf{Data Configuration}} \\
\hline
Training Set Size & 80\% of total events (240K)\\
Test Set Size & 20\% of total events (60k) \\
Validation Strategy & Hold-out test set for early stopping \\
Class Stratification & Enabled (preserves class ratios) \\
Random Seed & 42 (reproducibility) \\
Data per Tree & Full training set at each round \\
Feature Information & 95 kinematic, angular, jet flavor-tagging scores, and event shape variables \\
\hline\hline
\end{tabular}
}
\end{table}

\end{document}